\documentclass[twocolumn,twocolappendix,times,tighten,trackchanges]{aastex631} 
\usepackage[version=3]{mhchem}
\usepackage{comment}
\usepackage{multirow}

\newcommand{\mum}{$\rm \mu \mathrm{m}~$}

\newcommand{\OIII}{[\textsc{Oiii}]\,}

\newcommand{\HII}{H{\sc ii}~}
\newcommand{\CII}{[C{\sc ii}]\,}
\defcitealias{Popping:2017}{P17}
\defcitealias{Toyouchi:2025}{T25}

\newcommand{\dtmconst}{constant DTM}
\newcommand{\dtmP}{\citetalias{Popping:2017}}
\newcommand{\dtmTfast}{\citetalias{Toyouchi:2025}-fast}
\newcommand{\dtmTslow}{\citetalias{Toyouchi:2025}-slow}

\newcommand{\tacc}{$\tau_{\rm acc,0}$}

\usepackage{color}
\usepackage{makecell}

\newcommand{\notice}[1]{\textcolor{orange}{\bf #1}}

\shorttitle{Dust Observables at $z \geq 10$ galaxies}
\shortauthors{Nakazato et al.}

\newcommand{\figdir}{./}

\begin{document}

\title{Dust in Galaxies at $z=10-15$: Attenuation and Re-emission Predictions from the FirstLight Simulations}

\correspondingauthor{Yurina Nakazato}
\email{ynakazato@flatironinstitute.org}
\author[0000-0002-0984-7713]{Yurina Nakazato}
\affiliation{Center for Computational Astrophysics, Flatiron Institute, 162 5th Avenue, New York, NY 10010}
\author[0000-0002-7779-8677]{Akio K. Inoue}
\affiliation{Waseda Research Institute for Science and Engineering, Faculty of Science and Engineering, Waseda University, 3-4-1 Okubo, Shinjuku, Tokyo 169-8555, Japan}
\affiliation{Department of Physics, School of Advanced Science and Engineering, Faculty of Science and Engineering, Waseda University, 3-4-1, Okubo, Shinjuku, Tokyo 169-8555, Japan}
\author[0000-0002-6748-6821]{Rachel S. Somerville}
\affiliation{Center for Computational Astrophysics, Flatiron Institute, 162 5th Avenue, New York, NY 10010}
\author[0000-0003-3467-6079]{Daisuke Toyouchi}
\affiliation{Theoretical Astrophysics, Department of Earth and Space Science, The University of Osaka, 1-1 Machikaneyama, Toyonaka, Osaka
560-0043, Japan}
\author[0000-0002-5012-6707]{Kosei Matsumoto}
\affiliation{Sterrenkundig Observatorium Department of Physics and Astronomy Universiteit Gent, Krijgslaan 281 S9, B-9000 Gent, Belgium}
\author[0000-0002-8680-248X]{Daniel Ceverino}
\affiliation{Departamento de Fisica Teorica and CIAFF, Facultad de Ciencias, Universidad Autonoma de Madrid, Cantoblanco, 28099 Madrid, Spain}
\affiliation{CIAFF, Facultad de Ciencias, Universidad Autonoma de Madrid, 28049 Madrid, Spain}

\begin{abstract}
Dust masses and grain properties in $z>10$ galaxies remain poorly constrained, as JWST UV attenuation and ALMA dust re-emission are rarely modelled consistently. We perform post-processed 3D dust radiative transfer on 208 snapshots of 99 galaxies at $z=10$--15 from the FirstLight simulations ($M_*=10^{6.5}$--$10^{9.5}\,M_\odot$), applying twelve dust models per snapshot (2496 runs): four dust-to-metal (DTM) prescriptions and three grain size distributions. Metallicity-dependent DTM models reproduce ALMA dust-mass upper limits at $z\gtrsim10$ and detections at $z\sim7$, as dust-to-stellar mass ratios rise sharply near $Z\sim0.2\,Z_\odot$, typically at $M_*\gtrsim10^{8.5}\,M_\odot$, where grain growth by metal accretion becomes dominant. A constant DTM of 0.4 instead violates more than half of the upper limits. For dust attenuation, SMC-like dust is favored by the observed $M_{\rm UV}$--$\beta_{\rm UV}$ relation. The observed UV slopes ($\beta_{\rm UV}\simeq-2.3$) favor small grains ($a\lesssim\lambda_{\rm UV}/2\pi\simeq0.025\,\mu{\rm m}$) producing steep UV extinction, together with a low abundance of carbonaceous grains or PAHs responsible for the 2175 \AA\ bump, consistent with silicate-rich dust. Large-grain stellar dust ($a_0=0.1\,\mu{\rm m}$) instead yields nearly grey extinction and UV slopes close to the intrinsic value ($\beta_{\rm UV,0}\simeq-2.6$). SMC and MW attenuation curves flatten at $A_{\rm V}\gtrsim0.3$, showing that relatively flat attenuation does not necessarily imply large grains. For dust re-emission, dust peak temperatures reach 60--80 K, exceeding the 40--50 K commonly assumed in ALMA dust-mass estimates. Predicted IR continuum flux increases with UV luminosity; at current ALMA sensitivities (10--20 $\mu$Jy), detections generally require $M_{\rm UV}\lesssim-21.5$. Roman wide-field surveys may identify such rare UV-bright galaxies.
\end{abstract}
\section{Introduction} \label{sec:introduction}
The James Webb Space Telescope (JWST) has now spectroscopically confirmed more
than forty galaxies at $z\geq10$ \citep[][see the compilation of
\citealt{Roberts-Borsani:2026}]{Arrabal_Haro:2023, Curtis-Lake:2023}, with the
most distant reaching $z>14$ (GS-z14-0: \citealt{Carniani:2024}; MoM-z14:
\citealt{Naidu:2026}). Since JWST/NIRCam and NIRSpec trace the rest-frame UV at
$z>10$, they can constrain the dust attenuation in these galaxies. The
JWST-observed galaxies at $z>10$ have $-18.5 \lesssim M_{\rm UV} \lesssim -21.5$ \citep[see Figure 1 of][]{Roberts-Borsani:2026}, and
the abundance of the bright galaxies with $M_{\rm UV}\lesssim-20$ implies that the UV luminosity function exceeds pre-JWST theoretical
predictions at the bright end \citep[e.g.,][]{Harikane:2023, Finkelstein:2023, Adams:2024, Donnan:2024,
Castellano:2023, Leung:2023, Robertson:2024, Whitler:2025, Harikane:2025_UVLF}.

The dust content of these galaxies is an important factor in interpreting this excess of UV-bright galaxies, since dust attenuation directly affects their observed UV luminosities. Constraining their dust properties is therefore essential for distinguishing among the proposed explanations for the discrepancy between observations and theoretical predictions.

JWST can constrain the dust attenuation observables.
The UV slopes measured from NIRCam photometry and from spectroscopy
\citep[e.g.,][]{Topping:2022, Topping:2024, Cullen:2024, Morales:2024, Tang:2026_UVslope} are typically
blue, $\beta_{\rm UV}\simeq-2.3$, but with a large scatter ranging from $-3.0$
to $-1.5$. Indeed, in the compilation of \citet{Roberts-Borsani:2026}, 7 of 41
galaxies at $z\geq 10$ have \textit{red} slopes with $\beta_{\rm UV}>-2.0$, none of which
shows C\,\textsc{iv} emission in its prism spectrum, disfavoring AGN as the
origin of the reddening. Attenuation curves are also beginning to be constrained up to the Epoch of
Reionization using NIRCam grism or NIRSpec
\citep[e.g.,][]{Markov:2025, Shivaei:2026, Chworowsky:2026}, although
studies at $z\gtrsim10$ remain scarce owing to the lack of rest-optical data.

The Atacama Large Millimeter/submillimeter Array (ALMA), on the other hand, constrains the re-emission properties. Direct searches for the rest-frame infrared (IR) continuum have been carried out alongside
searches for FIR emission lines \citep[e.g.,][]{Algera:2026_z8_10_OIII,
Schouws:2025, Zavala:2024, Witstok:2026_ALMA}\footnote{GN-z11 was instead
observed with NOEMA, which also resulted in a non-detection of the dust
continuum \citep{Fudamoto:2024}.}, but all have resulted in non-detections at
$z>10$, yielding only upper limits (see the compilation of
\citealt{Bakx:2026}). Taken together, these observations indicate that galaxies
at $z>10$ are on average moderately blue and poor in dust.

At $z\sim6-8$, in contrast, ALMA has detected a considerable number of galaxies
hosting $M_{\rm dust}\sim10^{7}-10^{8}\,M_{\odot}$ of dust \citep{Riechers:2013, Watson:2015, Marrone:2018, Zavala:2018, Tamura:2019, Hashimoto:2019, Harikane:2020, Fudamoto:2021, Inami:2022, Bowler:2022, Witstok:2022, Mitsuhashi:2024_SERENADE, Umehata:2025}, with the most
distant dust-detected galaxy being MACS0416-Y1 at $z=8.3$ \citep{Tamura:2019, Bakx:2020, 
Bakx:2025}. The fact that galaxies are dust-poor at $z\geq10$ (cosmic time
$\sim400-500\,{\rm Myr}$) yet host such large dust reservoirs only
$\sim100-200\,{\rm Myr}$ later implies that a rapid build-up of dust took place
in between. This is the primary motivation for understanding the dust in the
most distant galaxies at $z\geq10$.

The impact of dust on the observed UV emission of galaxies at $z>10$ is primarily governed by two quantities: the dust mass and the grain properties (size, composition). In the UV, once the grain size becomes comparable to or larger than the wavelength ($2\pi a/\lambda \gtrsim 1$), the extinction efficiency $Q_{\rm ext}$ saturates at a value of order unity, and the mass extinction opacity therefore scales as $\kappa_{\rm UV} \propto Q_{\rm ext}/a \propto a^{-1}$; larger grains are less efficient absorbers per unit mass. At fixed dust mass, the UV extinction can thus vary by an order of magnitude depending on the grain size distribution. In addition, grains smaller than the wavelength give a much steeper wavelength dependence than large grains, which approach a grey extinction law, so the \textit{shape} of the attenuation curve is likewise sensitive to the size distribution. Measurements of UV slopes and attenuation curves with JWST therefore offer a promising route to constraining grain sizes and the dust-star geometries.

The UV radiation absorbed by dust is re-emitted in the rest-frame IR, making rest-frame FIR observations essential for constraining the total dust mass, although the inferred mass remains degenerate with the assumed dust temperature. The dust mass itself is set by the competition between stellar dust production and grain growth in the ISM. \citet{Asano:2013} showed analytically that there exists a critical metallicity $Z_{\rm crit}$ above which grain growth by metal accretion contributes more to the dust budget than production by core-collapse supernovae and AGB stars. This picture is supported by a number of analytical and semi-analytical models \citep{Popping:2017, Toyouchi:2025, Toyouchi:2026}, and is consistent with the observed dust-to-metal (or dust-to-gas) ratio versus metallicity relation in the local Universe \citep[e.g.,][]{Remy-Ruyer:2014}. Recent efforts have attempted to place similar constraints at $z > 6$ \citep[e.g.,][]{Algera:2026_REBELS}, although gas mass estimates at these redshifts carry substantial uncertainties; the ALMA Cycle~12 Large Program PHOENIX (PI: Schouws) is expected to improve this situation considerably. It is therefore essential to investigate how the dust observables; the attenuation and re-emission properties, respond jointly to changes in the dust mass and the grain size distribution, in order to interpret forthcoming JWST and ALMA constraints.


On the theoretical side, analytical and semi-analytical models incorporating
dust evolution (production, grain growth, and destruction) have been applied to galaxies at $z>6$ and broadly reproduce the observed evolution of the dust mass \citep[e.g.,][]{Popping:2017, Dayal:2022, Toyouchi:2025}. These models, however, do not account for the internal spatial structure of galaxies, or simply assume disk-like geometries, and therefore cannot predict the attenuation curve or dust temperature, both of which depend on the relative spatial distributions of stars and dust\footnote{\citet{Toyouchi:2026} do derive attenuation curves, but for an idealized disk galaxy with a prescribed distribution of ISM clumps and without accounting for three-dimensional scattering.}.

Cosmological simulations that include ``live'' dust models have also been developed \citep[e.g.,][]{Graziani:2020, Esmerian_Gnedin:2022, Choban:2025, Lewis:2023, Lower:2023, Jones:2024, Kannan:2022, Kannan:2025}. Recently, \citet{Narayanan:2026} and COLIBRE \citep{Trayford:2026, Vijayan:2026} have followed the evolution on the fly and examined the dust evolution over $z\sim 6-10$. These studies, however, have focused mainly on the evolution of the dust mass and of the grain sizes themselves, and none has systematically reported both the attenuation and the re-emission observables. A more fundamental limitation is that solving dust evolution on the fly yields only a single dust model per simulation, so that the effect of assuming different dust growth models or grain size distributions cannot be compared for the same galaxy sample.

In this work, we therefore take a complementary approach: we perform
post-processed three-dimensional dust radiative transfer (RT) on the
FirstLight cosmological zoom-in simulations \citep{Ceverino:2017, Ceverino:2026}. Since FirstLight does not include a live dust model, we instead systematically vary the assumed dust-to-metal ratio (DTM) and grain size distribution in post-processing. While previous dust-RT studies with FirstLight \citep{Mushtaq:2023, Nakazato:2026} focused mainly on $z\simeq6$--$9$ galaxies with a fixed dust-to-metal ratio and a single grain size distribution, here we target galaxies at $z\simeq10$--$15$ and explore the range of dust properties that are plausible at these redshifts. This approach allows us to apply several sets of dust properties to identical distributions of gas density, metallicity, and stars, so that the differences we find can be attributed to the assumed dust properties alone rather than to differences in galaxy evolution.

This paper is organized as follows. Section \ref{sec:method} describes the simulations and our radiative transfer calculations. Section \ref{sec:results} presents the scaling relations of the simulated galaxies and discusses how dust mass varies among different dust-to-metal ratio models. Sections \ref{sec:dust_attenuation_properties} and \ref{sec:dust_reemission_properties} show the resulting dust attenuation and reemission properties with different DTMs and grain size distributions. In Section \ref{sec:discussion} we discuss the most favorable combination of DTM and size distributions, and the detectability of the dust continuum with ALMA. We summarize our conclusions in
Section \ref{sec:summary}. 

\section{Method} \label{sec:method}
\subsection{Cosmological Zoom-in Simulations: FirstLight}  \label{subsec:FirstLight}
We use the cosmological zoom-in simulation suite FirstLight. The simulations used in this work were first presented in \citet{Ceverino:2017} and have already been used to study several high-redshift topics, including galaxy morphologies \citep{Ceverino:2021, Nakazato:2024}, star formation histories \citep{Ceverino:2018, Ceverino:2024, Dome:2024}, chemical evolution \citep{Langan:2020, Nakazato:2023}, UV/IR continuum emission \citep{Ceverino:2019, Mushtaq:2023, Nakazato:2026}, and size evolution \citep[]{Cataldi:2026, Ceverino:2026}. 

We use the simulation code ART, which employs Adaptive Mesh Refinement \citep[AMR;][]{Kravtsov:1997, Kravtsov:2003, Ceverino:2009}. 
The FirstLight simulations consist of four different simulation boxes (see Table 1 of \citealp{Ceverino:2026}), and here we use two with the side lengths of 40 and 80 comoving Mpc\,$h^{-1}$. We select all dark matter halos with a maximum circular velocity greater than $V_{\rm circ} = 180 (250)\, {\rm km\,s^{-1}}$ at $z=5$ for 40 (80) cMpc\,$h^{-1}$ box, resulting in a sample of 62 (53) halos. For each halo, we perform zoom-in simulations with a dark matter particle mass of $8 \times 10^4\, (6\times 10^5)\, M_\odot$ and a minimum stellar particle mass of $10^3\, (10^4) M_\odot$. All simulations were evolved down to $z = 5.25$. The smallest cell size achieved is $\Delta x_{\rm min}=15 \,(31)\, {\rm pc}\,(14/(1+z))$. To prevent artificial fragmentation at the smallest cell scale, a pressure floor is imposed to ensure that the Jeans length is always resolved by at least seven cells \citep[also see Appendix~B of \citealp{Nakazato:2024}]{Ceverino:2010}.

Stars form in high-density ($\rho > \rho_{\rm th}$) and low-temperature ($T < T_{\rm th}$) gas cells. In our simulations, we adopt thresholds of $\rho_{\rm th} = 0.035\, M_\odot\, {\rm pc^{-3}}$ ($n_{\rm th} = 1\,{\rm cm^{-3}}$) and $T_{\rm th} = 10^4\, {\rm K}$. We adopt a stochastic model in which the star formation probability scales with the gas free-fall time \citep{Ceverino:2009}. This prescription ensures that higher-density gas cells are more likely to form stars, resulting in the Kennicutt–Schmidt relation \citep{Kennicutt:1998}. Our simulations follow radiative cooling by atomic hydrogen and helium, metal ions and atoms, and molecular hydrogen calculated using \textsc{cloudy} \citep{Ferland:1998}. We also include photoionization heating by the redshift-dependent UV background with partial self-shielding \citep{Haardt_Madau:1996}.

In the FirstLight simulations, we incorporate the following stellar feedback mechanisms: supernovae (SNe) and stellar winds, radiation pressure by ionizing photons, 
and IR radiation pressure via a subgrid model. The first two are implemented as thermal pressure terms with a constant heating rate for 40 Myr, whereas the latter two are implemented as non-thermal pressure terms \citep{Ceverino:2014}. In addition to thermal feedback, we include the injection of momentum from the (unresolved) expansion of gaseous shells driven by SNe and stellar winds \citep{Ostriker_Shetty:2011}. Further details can be found in \citet{Ceverino:2017}.

The 115 galaxies in our sample span stellar masses of $\log_{10}(M_*/M_\odot) = 9.6-11$ at $z=5$. The maximum resolution in the zoom-in hydrodynamical simulations, particularly in dense clumps, enables us to resolve gas densities of $\sim 10^3\, {\rm cm^{-3}}$ with temperatures of $\sim 300\, {\rm K}$. Among these galaxies, snapshots are produced with a spacing of $\Delta a = 0.001$ in the cosmic expansion factor, starting once the halo mass exceeds $10^{10}\,M_\odot$, a threshold that allows well-resolved galaxies. Consequently, not all galaxies have snapshots extending back to $z=15$, and less massive haloes appear in the sample only at later times. We select the 99 galaxies that have at least one snapshot at $z\gtrsim9.5$, which provides a total of 208 snapshots for our analysis at $z=9.5-15$. The number of
snapshots decreases towards higher redshift, with $98$, $61$, $33$, $13$, $2$,
and $1$ snapshots at $z\simeq10$, $11$, $12$, $13$, $14$, and $15$,
respectively.

 \subsection{Post-Processing Radiative Transfer} \label{subsec:skirt}
Dust attenuation of all simulated galaxies (99 galaxies, 208 snapshots for $z=10, 11, 12, 13, 14, 15$) is calculated using the post-processing radiative transfer code SKIRT \citep[version 9, ][]{Baes:2011, Camps_Baes:2020}. SKIRT is an open-source, three-dimensional Monte Carlo dust RT code that self-consistently calculates dust temperatures and re-emission by accounting for both absorption and anisotropic scattering processes.
To perform the RT calculations, we specify the stellar and nebular emission properties (Section \ref{subsubsec:stellar_emission}), dust spatial distribution (Section \ref{subsubsec:dust_spatial_distribution}), and adopted dust composition and grain-size distributions (Section \ref{subsubsec:dust_size_composition}) as input parameters. Each simulation snapshot is processed within a cubic region with a side length of 0.3$R_{\rm vir}$, where $R_{\rm vir}$ is the virial radius of each halo. 

\subsubsection{Stellar Emission} \label{subsubsec:stellar_emission}
We construct the SEDs of individual stellar particles using binary stellar population models from \textsc{BPASS} \citep{Eldridge:2017}, based on the mass, age, and metallicity of each stellar particle. Each stellar particle is treated as a single starburst. Among the \textsc{BPASS} models, we adopt a Chabrier IMF with an upper mass limit of $300\, M_\odot$.

For stellar particles younger than 15 Myr, we reassign their ages following \citet{Nakazato:2023}. This adjustment is needed because our simulations form stellar particles with a fixed time step of $\Delta \tau_{\rm SF} = 5 \, {\rm Myr}$\footnote{The time step width in the simulation is much shorter than this, typically 1000 yr.}, while the simulation output times are not synchronized with this interval. The reassignment accounts for continuous star formation within each discrete timestep.

We include nebular continuum emission in our RT calculations. This is motivated by recent JWST observations suggesting that nebular emission may dominate the UV continuum in some high-redshift star-forming systems \citep[e.g.,][]{Cameron:2024, Yanagisawa:2025, Reumert:2026}, and by several theoretical studies that have investigated its effect on the UV slopes \citep[e.g.,][]{Inoue:2011, Ceverino:2019, Katz:2025, Narayanan:2025}. We select stellar particles younger than 10 Myr and assume that each of them is surrounded by an \HII region. The electron density is set to the gas number density of the cell hosting the stellar particle, and the electron temperature is assumed to be
$T_{\rm e} = 1.5 \times 10^4\, {\rm K}$, which is a typical value for $z>6$ observed galaxies obtained from the \OIII auroral line \citep[e.g.,][]{Nakajima:2023, Morishita:2024}. We have confirmed that our results are insensitive to this choice: adopting $T_{\rm e} = 1.0\ (2.0) \times 10^4\, {\rm K}$ changes the value of the intrinsic UV slopes by only 0.01, as shown in Appendix \ref{sec:nebular_continuum_contribution}. For the volume of each \HII region, we adopt $V_{\rm HII} = {\rm min}(V(R_{\rm s}), V_{\rm cell})$, where $V(R_{\rm s})$ is the volume of a Str\"omgren sphere. We consider only the free-free, free-bound, and two-photon continua from hydrogen, since the contribution of helium is negligible \citep{Inoue:2010}. The nebular continuum has been newly implemented in SKIRT (Matsumoto et al. in prep.), and we have verified its consistency by comparing the intrinsic nebular continuum spectra computed with SKIRT against analytical calculations following \citet{Inoue:2010}.

We first studied the change in the UV slope induced by the nebular continuum without dust attenuation, $\delta \beta_{\rm neb, 0}\equiv \beta_{\rm (stellar+nebular), 0} - \beta_{\rm stellar, 0}$. For an electron temperature of $T_{\rm e} = 1.5\times 10^4\, {\rm K}$, we find that $\delta \beta_{\rm neb, 0}$ has a median value of 0.044 with maximum values up to $\delta \beta_{\rm neb, 0} = 0.3$ (see the right panel of Figure \ref{fig:nebular_contribution}). After the post-processing RT, however, the slope difference induced by the nebular continuum decreases to $\delta \beta_{\rm neb} = 0.015$--$0.021$ (median values), depending on the DTM models. This is because the nebular continuum originates from star-forming regions where the dust column density is high, and it is therefore attenuated more strongly than the total stellar continuum.
Since dust attenuation can change the slope by $\Delta \beta \equiv \beta_{\rm UV}- \beta_{\rm UV,0} = 0.2$--$1.0$ (see details in Section~\ref{subsec:UVslope_vs_MUV}), we thus conclude that the nebular continuum has only a minor effect on the UV slope in a
statistical sense. Further details are given in Appendix~\ref{sec:nebular_continuum_contribution}.

The SED (including both stellar and dust emission) is sampled using 150 logarithmically spaced wavelength bins, covering the range from 0.1 to 1000\,\mum. To ensure good convergence, a total of $10^7$ photon packets are emitted from the stars in each calculation. The numerical convergence with respect to the number of photon packets and the wavelength binning is verified in Appendix~A of \citet{Nakazato:2026}. 

\begin{figure}
    \centering
    \includegraphics[width =\linewidth, clip]{\figdir/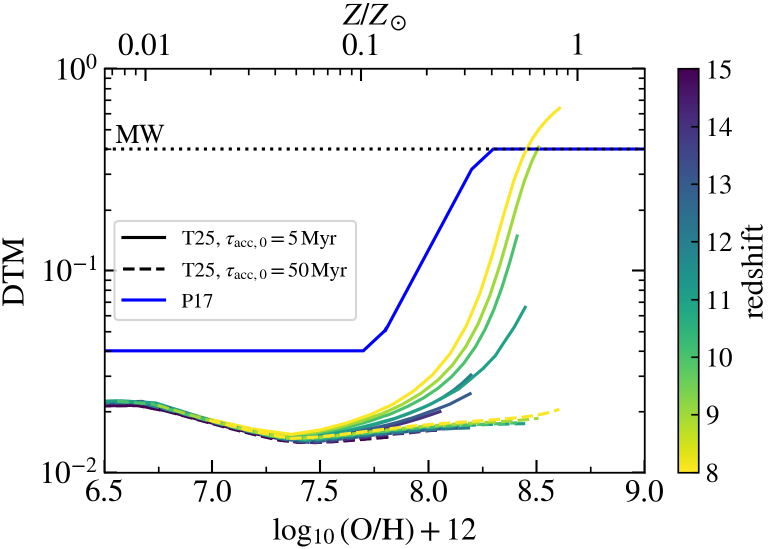}
    \caption{Dust-to-metal mass ratio (DTM) as a function of gas-phase metallicity. The black dashed line shows the constant model with the MW value ${\rm DTM}_{\rm MW}=0.4$. The blue line is the variable DTM model from \citet{Popping:2017}. The colored lines show the variable DTM models from \citet{Toyouchi:2025}, with a rapid growth scenario ($\tau_{\rm acc, 0} = 5\,{\rm Myr}$, solid) and a slow growth scenario ($\tau_{\rm acc, 0} = 50\,{\rm Myr}$, dashed).}
    \label{fig:DTM_Z}
\end{figure}

\subsubsection{Dust Spatial Distribution} \label{subsubsec:dust_spatial_distribution}
The amount of dust in each gas cell is taken from the hydrodynamical simulation, assuming that dust traces metals according to the dust-to-metal ratio ${\rm DTM} \equiv M_{\rm dust}/M_{\rm metal}$\footnote{The metal mass here contains all the metals in gas-phases and dust.}. Since a large number of observed $z \gtrsim 10$ galaxies are expected to have metallicities below the critical value at which dust accretion growth becomes efficient, we consider four DTM prescriptions: a fixed DTM and three metallicity-dependent variable DTM models as shown in Figure \ref{fig:DTM_Z}.

For the fixed DTM model, we adopt ${\rm DTM} = 0.4$, consistent with the Milky Way value \citep{Zubko:2004}. For the variable DTM, we adopt the metallicity-dependent DTM predicted by two models of dust growth via metal accretion, \citet{Popping:2017} and \citet{Toyouchi:2025} (hereafter P17 and T25).

\citetalias{Popping:2017} follow dust production, growth, and destruction within a semi-analytic model of galaxy formation \citep{Somerville:2008}. Their fiducial model adopts a dust growth timescale of $\tau_{\rm acc,0} = 15\,{\rm Myr}$ at solar metallicity $Z = Z_\odot$\footnote{We adopt $Z_\odot = 0.0142$ and $\log_{10}({\rm O/H})_\odot + 12 = 8.69$ throughout this paper \citep{Asplund:2009}.}, and the DTM rises toward the Milky Way value at $Z \sim 0.2\,Z_\odot$. Since their predicted DTM--metallicity relation shows
only a weak redshift dependence over $z = 6-9$ \citepalias[see Figure 5 of][]{Popping:2017}, we adopt a single fitting function at all redshifts:
\begin{equation}
    \log_{10}\left(\frac{\rm DTM}{\rm DTM_\odot}\right) =
    \begin{cases}
        -1 & (x \leq 7.75) \\
        2x - 16.5 & (7.75 < x < 8.25) \\
        0 & (x \geq 8.25),
    \end{cases}
    \label{eq:DTM_P17}
\end{equation}
where $x \equiv 12 + \log_{10}({\rm O/H})$ is the gas-phase metallicity and
${\rm DTM_\odot} = 0.4$.

\citetalias{Toyouchi:2025} analytically model galaxy evolution including dust
growth via metal accretion with a timescale
$\tau_{\rm acc} = \tau_{\rm acc,0}(Z/Z_\odot)^{-1}$, where $\tau_{\rm acc,0}$ is the
metal accretion timescale onto dust grains at solar metallicity. They consider two cases: rapid growth ($\tau_{\rm acc,0} = 5\,{\rm Myr}$) and delayed growth ($\tau_{\rm acc,0} = 50\,{\rm Myr}$), yielding different DTM--metallicity relations as shown in Figure~\ref{fig:DTM_Z}. In the rapid growth case, DTM increases sharply toward ${\rm DTM} \sim 0.4$ as the gas-phase metallicity reaches $12 + \log({\rm O/H}) \sim 8.0$, driven by efficient dust growth via metal accretion \citep{Inoue:2011_dust_growth, Asano:2013}. A mild redshift dependence is also present at $12 + \log({\rm O/H}) \gtrsim 8.0$, reflecting the redshift dependence of metal enrichment timescales. In contrast, the delayed growth model shows little evolution in DTM with either metallicity or redshift, as the long accretion timescale suppresses dust growth, particularly at high redshift. For example, a galaxy with $Z_{\rm gas} = 0.1\,Z_\odot$ has $\tau_{\rm acc} = 500\,{\rm Myr}$, comparable to the Hubble time at $z \gtrsim 10$, leaving insufficient time for significant dust growth via accretion.

The $\tau_{\rm acc,0}$ adopted in \citetalias{Popping:2017} is intermediate between the two cases of \citetalias{Toyouchi:2025}, while the metallicity at which the DTM rises sharply is similar to that of the rapid-growth model. Below this critical metallicity, the higher DTM in  \dtmP \,mainly reflects its larger adopted stellar dust yield ($y_{\rm d}=0.15$) compared to \citetalias{Toyouchi:2025} ($y_{\rm d}=0.1$). This higher initial dust abundance may partly compensate for the longer accretion timescale in \dtmP, since the dust growth rate via metal accretion is proportional to the pre-existing dust mass, $\dot{M}_{\rm d,acc}\propto M_{\rm d}/\tau_{\rm acc}$. As a result, the two models exhibit a similar onset of rapid dust growth, although the subsequent DTM increase is steeper in the rapid growth model of \citetalias{Toyouchi:2025}.

We apply these DTM prescriptions cell-by-cell, using the gas-phase
metallicity of each cell to compute the local dust mass; for the variable
DTM models (\citetalias{Popping:2017} and \citetalias{Toyouchi:2025}), the
DTM itself also depends on this cell metallicity, and for
\citetalias{Toyouchi:2025} additionally on the redshift of each snapshot.

\begin{figure*}
    \centering
    \begin{minipage}{0.49\linewidth}
        \centering
        \includegraphics[width=\linewidth, clip]{\figdir/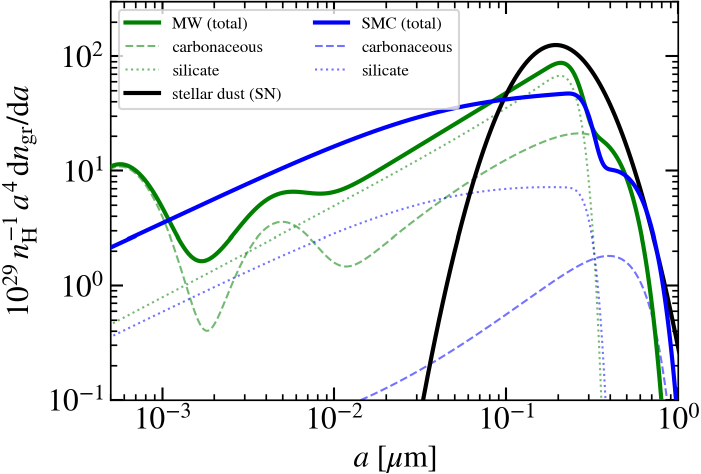}
    \end{minipage}
    \hfill
    \begin{minipage}{0.49\linewidth}
        \centering
        \includegraphics[width=\linewidth, clip]{\figdir/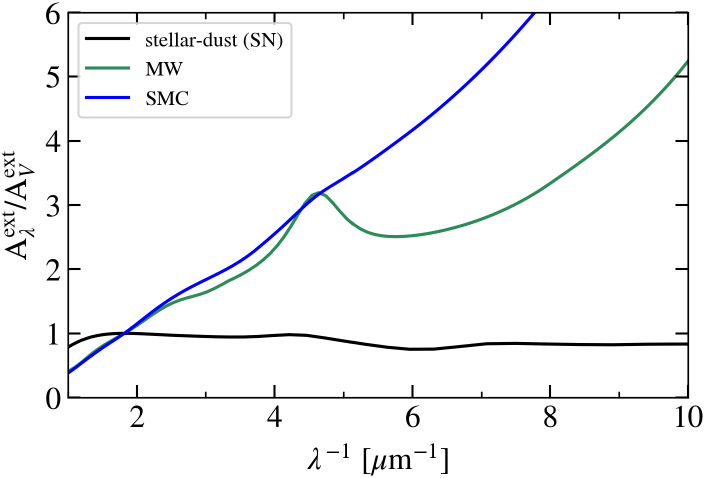}
    \end{minipage}
    \caption{Left: The normalized grain size distribution for SMC (blue), MW (green), and stellar dust (black) used in our dust RT calculations. The SMC and MW cases are calculated based on \citet{Weingartner_Draine:2001} with carbonaceous (dashed) and silicate grains (dotted). Right: Dust extinction curves for SMC \citep{Gordon:2003}, MW \citep[green;][]{Cardelli:1989}, and stellar dust (black). 
    }
    \label{fig:size_extinction_SN}
\end{figure*}

\begin{figure}
    \centering
    \includegraphics[width =\linewidth, clip]{\figdir/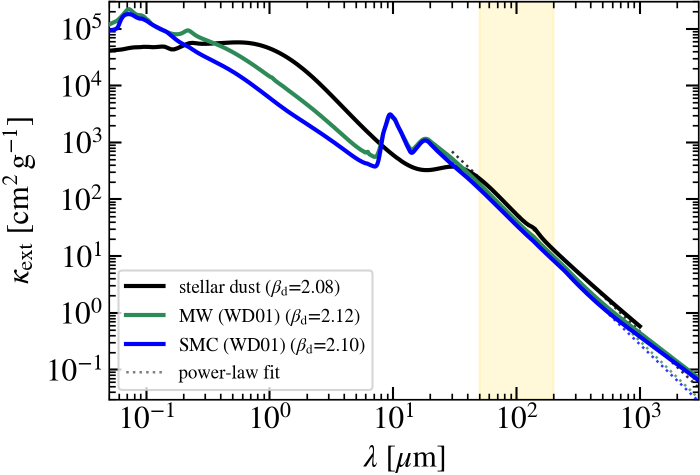}
    \caption{The extinction opacity per unit dust mass as a function of wavelength. The solid lines follow the same color scheme as in
Figure~\ref{fig:size_extinction_SN}. The dotted lines show power-law fits over
the rest-frame FIR range (shaded in yellow, $50$--$200\,\mu$m) used to derive
the dust emissivity index $\beta_{\rm d}$, whose best-fit values are given in
the legend.}
    \label{fig:kappa_ext_beta_d}
\end{figure}

\subsection{Dust Size Distribution and Composition
} \label{subsubsec:dust_size_composition}
The size distribution and composition of dust grains affect their absorption and scattering properties \citep[e.g.,][]{Draine:2014}. Since the exact nature of dust grains in high-redshift galaxies remains an open question, we consider three dust models: SMC-like, MW-like dust, and stellar dust driven by supernovae (SNe).

We adopt the \citet{Weingartner_Draine:2001} dust model and sample the
grain size distribution using 10/10/0 bins (graphite/silicate/PAH) for
SMC-like dust and 10/10/10 bins for MW-like dust, the latter additionally
including PAHs (polycyclic aromatic hydrocarbons). We include SMC-like
dust because small-grain-dominated size distributions, similar to that of
the SMC, are expected to be characteristic of the low-metallicity
environments typical of high-redshift galaxies \citep[0.15-0.2 $Z_\odot$;][]{Choudhury:2018, Reddy:2018, Liang:2021, Vijayan:2022}. For MW dust, our motivation comes from recent JWST observations that have investigated
dust attenuation curves at $z > 4$
\citep{Fisher:2025, Ormerod:2025, Witstok:2023, Markov:2023, Markov:2025},
finding that 20--25\% of galaxies in the REBELS and JADES samples show
evidence for a 2175\,\AA\ bump. This feature may originate from
carbonaceous grains including PAHs \citep{Li:2024, Lin:2025, Nanni:2025},
as seen in the MW extinction curve \citep{Cardelli:1989} but absent in
that of the SMC \citep{Gordon:2003}.

For stellar dust, freshly condensed grains from supernovae are theoretically expected to dominate the dust budget in $z > 10$ galaxies \citep{Schneider_Maiolino:2024, Narayanan:2026}. We adopt a log-normal size distribution following \citet{Hirashita_Aoyama:2019},
\begin{equation}
    \phi(a) = C_\phi \frac{1}{a} \exp \left[ - \frac{\ln^2(a/a_0)}{2\sigma^2} \right],
\end{equation}
where $a_0$ is the mean grain radius, $\sigma$ is the standard deviation, and $C_\phi$ is a normalization constant. We adopt $\sigma = 0.47$ and $a_0 = 0.1\, \mu {\rm m}$ following \citet{Asano:2013_grain_size, Hirashita_Aoyama:2019}. The size distribution of grains supplied by SNeII is biased to large ($\sim 0.1\, \mu {\rm m}$) grains due to the destruction of small grains by the reverse shock \citep[][but see also \citealt{Otaki:2026}]{Nozawa:2007, Bianchi_Schneider:2007}.
We consider only carbonaceous grains for SN dust, as \citet{Otaki:2026_size} show that carbonaceous dust dominates the grain composition for Type~II SNe.

The left panel of Figure~\ref{fig:size_extinction_SN} shows the size distributions for SMC (blue), MW (green), and stellar dust (black), where the SMC and MW distributions are calculated based on \citet{Weingartner_Draine:2001}\footnote{We plot the case with $R_V = 3.1$ and $b_{\rm C}=6\times 10^{-5}$, where $b_{\rm C}$ is the total C abundance per H nucleus in the log-normal populations 
}. 
We see that, in contrast to stellar dust, the MW size distribution has a large amount of small grains ($a \lesssim 0.1\, \mu{\rm m}$) with a power-law slope of $a^{-3.5}$ \citep[c.f., MRN;][]{Mathis:1977} and two log-normal components centered at $a_0 = 3.5$\,\AA\ and $30$\,\AA, which are associated with PAHs.

The right panel of Figure~\ref{fig:size_extinction_SN} shows the corresponding extinction curves. The stellar dust extinction curve is calculated using the optical properties from \citet{Draine_Lee:1984} and \citet{Laor_Draine:1993}, yielding a gray extinction curve with $A_{\rm UV}^{\rm ext}/A_{\rm V}^{\rm ext}=0.79$. In contrast, the SMC and MW extinction curves have $A_{\rm UV}^{\rm ext}/A_{\rm V}^{\rm ext}=4.77$ and $2.66$, respectively \citep{Cardelli:1989, Gordon:2003}. The MW extinction curve also exhibits a prominent UV bump at $\lambda^{-1}\simeq 4.6\,\mu{\rm m}^{-1}$ (i.e., $\lambda=2175$ \AA), commonly attributed to small carbonaceous grains, including PAHs. The steeper UV extinction curve of the SMC dust model reflects its grain size distribution and composition, particularly the contribution of small, predominantly silicate grains ($a\lesssim\lambda_{\rm UV}/2\pi\simeq 0.025\,\mu{\rm m}$) to wavelength-dependent UV extinction.

Figure~\ref{fig:kappa_ext_beta_d} shows the extinction opacity per unit dust mass for each grain model. The MW and SMC models have nearly identical opacities at $1600$\,\AA, whereas the stellar dust is lower by a factor of two, reflecting its larger characteristic grain size. We derive the dust emissivity index by fitting a power law over the
rest-frame 50--200~$\mu$m range, and find values close to $\beta_{\rm d}\simeq2$
for all three models ($\beta_{\rm d} = 2.08$--$2.12$). We therefore adopt
$\beta_{\rm d} = 2$ throughout this paper for simplicity. Table \ref{table:dust_models} summarizes the adopted dust model parameters and defines the corresponding model labels, yielding 12 dust models for each simulation snapshot. We note that the dust evolution timescale (i.e., the dust accretion timescale) depends on grain size (see also the caveats in
Section~\ref{subsec:caveats}), so the combinations of stellar dust with \dtmP\ or \dtmTfast\ are not necessarily self-consistent. However, the goal of this paper is to isolate the effects of the DTM model and the grain size distribution separately, and we therefore include this combination in our grid for that purpose.

\begin{table}[]
\centering
\caption{Summary of dust models used in this study.}
\begin{tabular}{llll}
\hline\hline
Name & Dust type & DTM models & \tacc \\
\hline
\multicolumn{4}{l}{\textit{SMC dust: \citet{Weingartner_Draine:2001} size distribution}} \\
\hline
SMC/constant  & SMC & 0.4 (fixed, MW value)          & --- \\
SMC/P17       & SMC & variable with $Z_{\rm gas}$ \citepalias{Popping:2017}      & 15 Myr \\
SMC/T25-fast  & SMC & variable with $Z_{\rm gas}$, $z$ \citepalias{Toyouchi:2025} & 5 Myr  \\
SMC/T25-slow  & SMC & variable with $Z_{\rm gas}$, $z$ \citepalias{Toyouchi:2025} & 50 Myr \\

\hline
\multicolumn{4}{l}{\textit{MW dust: \citet{Weingartner_Draine:2001} size distribution}} \\
\hline
MW/constant  & MW & 0.4 (fixed, MW value)          & --- \\
MW/P17       & MW & variable with $Z_{\rm gas}$ \citepalias{Popping:2017}      & 15 Myr \\
MW/T25-fast  & MW & variable with $Z_{\rm gas}$, $z$ \citepalias{Toyouchi:2025} & 5 Myr  \\
MW/T25-slow  & MW & variable with $Z_{\rm gas}$, $z$ \citepalias{Toyouchi:2025} & 50 Myr \\
\hline
\multicolumn{4}{l}{\textit{stellar dust: log-normal \citep{Hirashita_Aoyama:2019}, $a_0=0.1\,\mu$m}} \\
\hline
stellar/constant  & stellar & 0.4 (fixed, MW value)          & --- \\
stellar/P17       & stellar & variable with $Z_{\rm gas}$ \citepalias{Popping:2017}      & 15 Myr \\
stellar/T25-fast  & stellar & variable with $Z_{\rm gas}$, $z$ \citepalias{Toyouchi:2025} & 5 Myr  \\
stellar/T25-slow  & stellar & variable with $Z_{\rm gas}$, $z$ \citepalias{Toyouchi:2025} & 50 Myr \\
\hline
\end{tabular}\label{table:dust_models}
\end{table}
\subsection{Other Settings}
We account for several additional physical effects in the radiative transfer calculations.
First, we include self-absorption of dust emission, which becomes important in optically thick, dense clouds.
Second, we include the transient heating function to compute non-local thermal equilibrium (NLTE) dust emission from stochastically heated small grains and PAH molecules \citep{Camps_Baes:2015}.
The effects of self-absorption and NLTE settings are discussed in Appendix~A of \citet{Nakazato:2026} \citep[see also][]{Behrens:2018}.
Third, we include dust heating by the cosmic microwave background (CMB), whose temperature increases with redshift as $T_{\rm CMB}(z) = (1+z)\,T_{{\rm CMB}(z=0)}$, making it non-negligible at high redshifts \citep{da_Cunha:2015}.

The radiative transfer calculations are iterated until the IR re-emission luminosity converges to within 3\%, or until a maximum of seven iterations is reached\footnote{The luminosity change between the 6th and 7th iterations of the
SKIRT self-consistent dust heating calculation has a median value of 5.97\% (16th--84th percentile: 3.94--8.67\% across our sample). We have confirmed that this level of non-convergence does not affect our results, as also noted in \citet{Nakazato:2026}.}.
The spatial grid of the dust medium is constructed using an octree structure, in which cells are recursively subdivided until each cell contains less than $10^{-6}$ of the total dust mass. The finest grid level corresponds to a physical cell width of ${\sim}4.9\,{\rm pc}$, approximately one-quarter of the minimum cell size in the hydrodynamical simulation. We have verified the accuracy of the grid discretization by confirming that the input and grid-discretized dust masses agree to within 0.09\%, and that the optical depth at rest-frame 5500\,\AA\ agrees to within 0.2\% along all three coordinate axes.
\begin{figure}
    \centering
    \includegraphics[width =\linewidth, clip]{\figdir/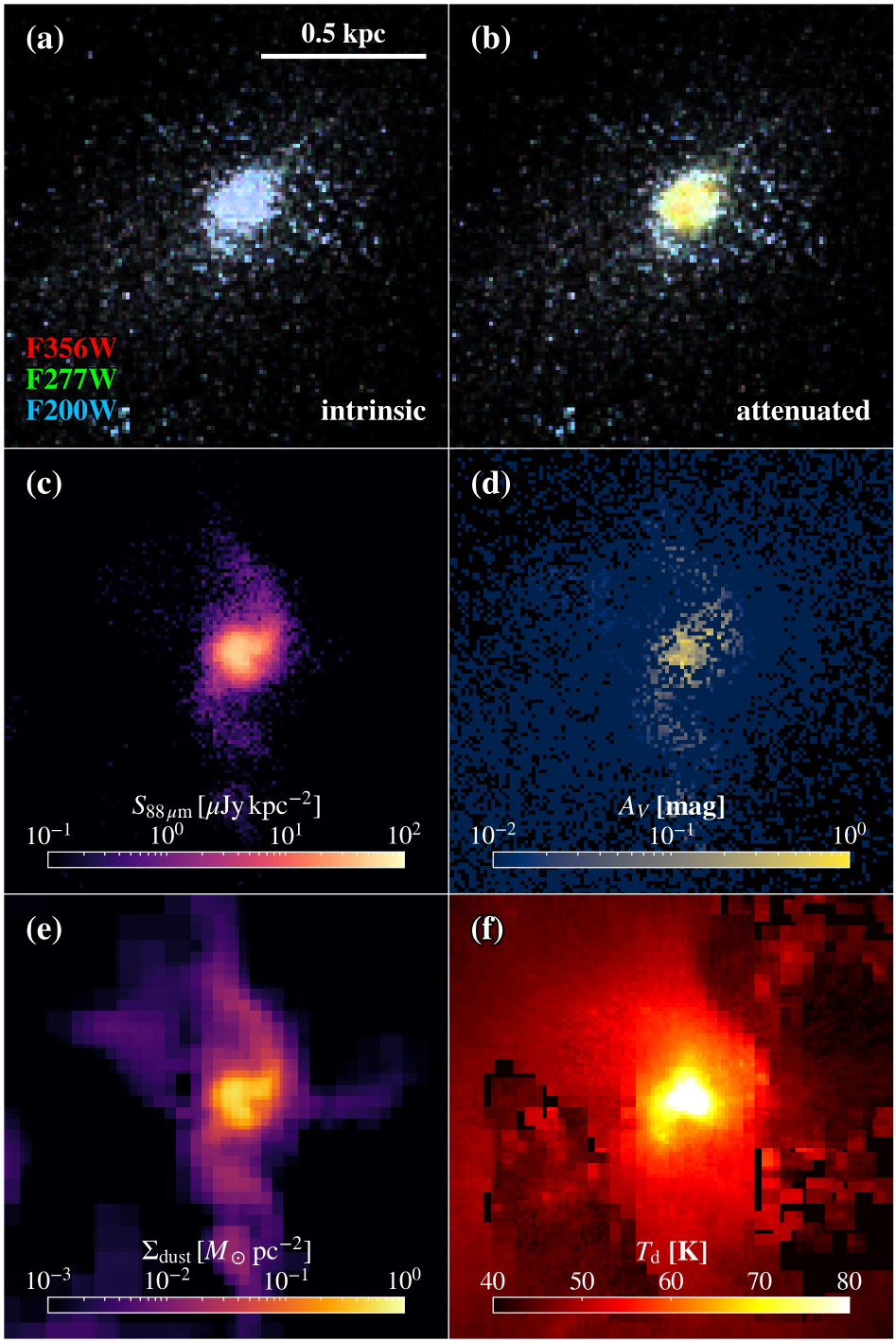}
    \caption{Projected distribution of one of our simulated galaxies, FL962 at $z=11$, using the \dtmTfast\, DTM model and SMC-like dust. (a) Mock JWST/NIRCam three-color image (F200W, F277W, F356W) without dust attenuation and (b) the same image with dust attenuation. (c) Dust continuum surface brightness at rest-frame $88\,\mu$m (no line emission), corresponding to ALMA Band~7 at this redshift. (d) $V$-band attenuation $A_{\rm V}$. (e) Dust column density and (f) mass-weighted dust temperature. Each panel shows a region of $1.35\,{\rm kpc}$ on a side, with a projected depth of $0.3R_{\rm vir}=2.25\, {\rm kpc}$.}
    \label{fig:FL962_a0p083_projection}
\end{figure}
\begin{figure*}
    \centering
    \includegraphics[width =\linewidth, clip]{\figdir/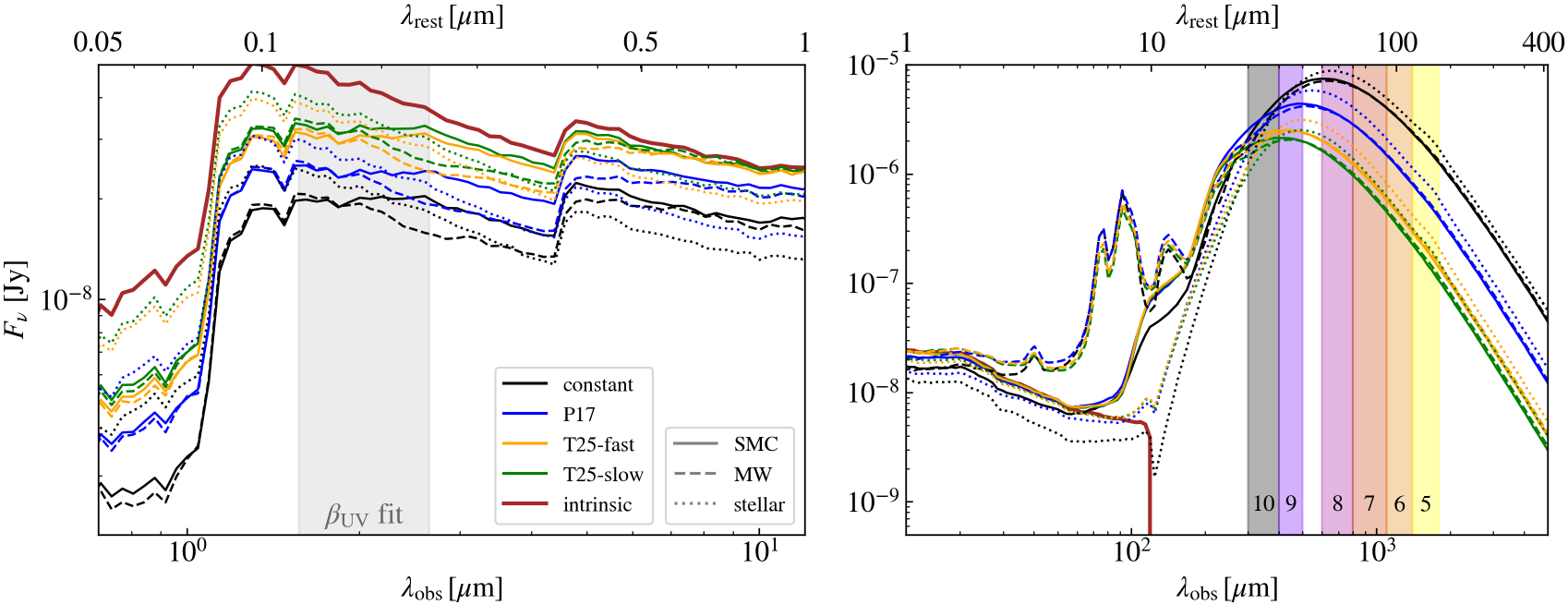}
    \caption{SEDs of the simulated galaxy FL962 at $z=11$, the same object as in Figure \ref{fig:FL962_a0p083_projection}. The left panel shows the rest-frame UV to optical range and the right panel the rest-frame infrared. The dark red solid line shows the intrinsic SED without dust attenuation, and the colored lines show the dust-attenuated SEDs for different DTM models: \dtmconst (DTM$=$0.4, black), \dtmP (blue), \dtmTfast (orange), and \dtmTslow (green). The solid, dashed, and dotted lines correspond to the SMC, MW, and stellar dust types, respectively. The grey shaded region in the left panel marks the wavelength range used to measure $\beta_{\rm UV}$, and the colored vertical bands in the right panel indicate the ALMA Bands 5 to 10.}
    \label{fig:SED_xy_neb_all_models}
\end{figure*}

\subsection{RT Results for a Representative Galaxy: FL962 at $z=11$}
Before presenting the statistical trends across our sample, we first
illustrate the RT results for a single example galaxy, FL962 at $z = 11.05$, one of our $z \geq 10$ galaxies from the FirstLight simulations. The galaxy has a stellar mass of $M_* = 2.7 \times 10^{8}\, M_\odot$, a gas-phase metallicity of $Z_{\rm gas} = 0.16\, Z_\odot$, a SFR averaged over the past 10~Myr of ${\rm SFR_{10}} = 3.7\, M_\odot\, {\rm yr^{-1}}$, comparable to the brighter end of spectroscopically confirmed $z \gtrsim 10$ galaxies \citep{Roberts-Borsani:2026}, and a compact stellar half-mass radius of
$r_{50,\star} = 360$~pc.

Figure~\ref{fig:FL962_a0p083_projection} shows the projected maps. All the panels other than panel (e) are obtained from the RT calculations, while panel (e) shows the input dust distribution. Here we adopt the \dtmTfast\ DTM model (rapid growth) and SMC-like dust, which yields a dust mass of $M_{\rm d} = 1.3 \times 10^{4}\, M_\odot$, with a corresponding dust-to-stellar mass ratio of only $M_{\rm d}/M_* = 4.8 \times 10^{-5}$, two orders of magnitude below the values inferred for $z \sim 7$ galaxies \citep[][see also Sect.~\ref{subsec:DtS_redshift}]{Algera:2026_REBELS}.

In a comparison of the mock images in panels (a) and (b), the central region becomes redder once dust attenuation is applied, since the shorter-wavelength bands are more strongly attenuated. This is also reflected in the wavelength dependence of the half-light radius: the intrinsic (attenuated) values in F200W ($\lambda_{\rm rest}\simeq1660$\,\AA), F277W ($\simeq2300$\,\AA), and F356W ($\simeq3000$\,\AA) are 259 (354), 259 (291), and 258 (276)~pc, respectively. In F200W the attenuated half-light radius is 37\% larger than the intrinsic one, because the attenuation is concentrated in the center and pushes the light-weighted radius outwards. The same central attenuation appears in the $V$-band attenuation map in panel (d), where the resolved values reach up to $A_V = 1.0$~mag even though the integrated SED gives only $A_V = 0.05$~mag, implying the integrated value is weighted towards the brighter, less obscured regions outside the center.

The absorbed UV radiation is re-emitted as rest-frame FIR continuum. Panel (c) shows the rest-frame $88\,\mu$m dust continuum (excluding emission lines), with a total flux of $S_{88 \mu{\rm m}} = 0.67\,\mu$Jy. Its distribution broadly follows the dust column density in panel (e), although it is more centrally concentrated with a half-light radius of 68.2 pc. Panel (f) shows the mass-weighted dust temperature $T_{\rm d,mass}$, which is higher in the inner regions, as expected from radiative equilibrium: $T_{\rm d} \propto u_\star^{1/(4+\beta_{\rm d})} \propto \left(L_\star/r^{2}\right)^{1/(4+\beta_{\rm d})}$, where $u_\star$ is the stellar radiation energy density, $L_\star$ the luminosity of the sources, $r$ the distance between the source and the dust grains, and $\beta_{\rm d}$ the dust emissivity index.

Figure~\ref{fig:SED_xy_neb_all_models} shows the SEDs of the same galaxy for the four DTM models and the three dust types. The DTM models yield dust masses of $M_{\rm d} = 2.2\times10^{5}$, $9.6\times10^{4}$, $1.3\times10^{4}$, and $9.2\times10^{3}\,M_\odot$ for \dtmconst, \dtmP, \dtmTfast, and \dtmTslow, respectively, i.e., a spread of 1.4 dex for a single galaxy. These differences translate into different attenuation in
the rest-frame UV and optical (i.e. different attenuation curves) and into IR flux densities that differ by $\sim$1 dex across ALMA Bands 5--8. Furthermore, the dust type changes the contribution of small carbonaceous grains, which is apparent in the rest-frame mid-IR where MW dust shows PAH features that are absent for stellar dust. These differences in the small-grain contribution affect both the IR peak temperature and the UV slope. Using these outputs for all snapshots, we discuss the dust observables statistically in Section~\ref{sec:dust_attenuation_properties} and \ref{sec:dust_reemission_properties}.

\section{Results} \label{sec:results}
\subsection{Scaling Relations}
\begin{figure}
    \centering
    \includegraphics[width =\linewidth, clip]{\figdir/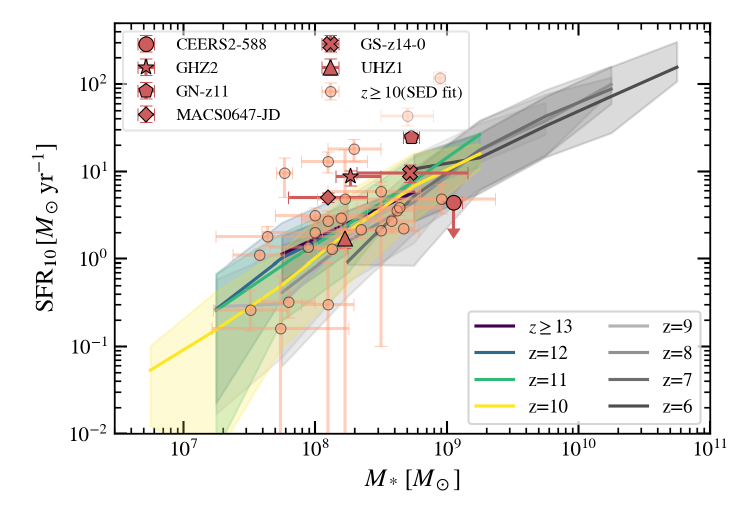}
    \caption{SFR versus stellar mass from $z=15$ to $z=10$. The solid lines show the median and the colored bands indicate the sample dispersion in the range of $5\%$--$95\%$. The ${\rm SFR}_{10}$ is calculated by summing the stellar mass formed over the past 10~Myr and dividing by 10~Myr. As a reference, the results for the simulated samples at $z=6$--$9$ are also shown as gray lines. The red plots show galaxies with H$\alpha$ detections from MIRI spectroscopy, with SFRs derived from H$\alpha$ luminosities.
The blue symbols correspond to: CEERS2-588 ($z=11.04$; \citealt{Harikane:2026}), GHZ2 ($z=12.33$; \citealt{Zavala:2025}), 
GN-z11 ($z=10.60$; \citealt{Alvarez-Marquez:2025}), MACS0647-JD ($z=10.17$; \citealt{Hsiao:2024}), GS-z14-0 ($z=14.18$; \citealt{Helton:2026_MIRI1}), and UHZ1 ($z=10.05$; \citealt{Alvarez-Marquez:2026}). The orange symbols with errorbars are the other 26 galaxies spectroscopically observed at $z\gtrsim 10$ whose SFRs are derived from SED fitting, for which the data are taken from Table~1 of \citet{Ferrara:2026}. }
    \label{fig:Ms_SFR_z10_15}
\end{figure}

We begin by presenting scaling relations of key physical properties of our simulated galaxies. Figure~\ref{fig:Ms_SFR_z10_15} shows the stellar mass--SFR relation at $z=10-13$, where the SFR is averaged over the past 10\,Myr. Our sample spans a stellar mass range of $M_* \simeq 5\times10^6$--$10^9\,M_\odot$ and a SFR range of $10^{-2}$--$30\,M_\odot\,{\rm yr}^{-1}$, enabled by our large-volume simulation boxes of 40 and 80\,cMpc\,$h^{-1}$. For comparison, we plot the $z \geq 10$ galaxies observed by JWST. The orange plots are compiled from \citet{Ferrara:2026}, where the SFRs are derived from SED fitting. The blue plots show galaxies with H$\alpha$ detections from MIRI spectroscopy \citep{Harikane:2026, Zavala:2025, Alvarez-Marquez:2025, Hsiao:2024, Helton:2026_MIRI1, Alvarez-Marquez:2026}, with SFRs derived from H$\alpha$ luminosities\footnote{U37126 at $z=10.255$ is observed by MIRI spectroscopy \citep{Marques-Chaves:2026}, but no nebular emission lines are detected, and we do not include this target.}. We see that the stellar mass range of the observed $z \geq 10$ galaxies is in good agreement with our galaxy samples, demonstrating that our simulated sample can be directly compared with current high-redshift observations.

\begin{figure}
    \centering
    \includegraphics[width =\linewidth, clip]{\figdir/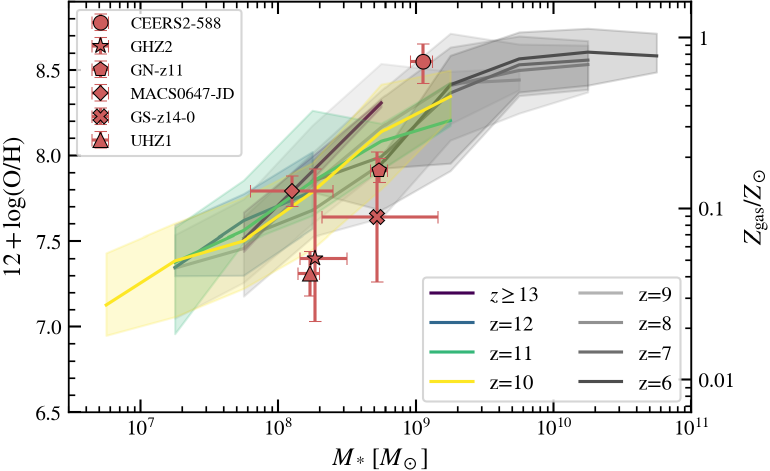}
    \caption{Gas-phase metallicity versus stellar mass for simulated galaxies from $z=15$ to $z=10$. The solid lines show the median and the colored bands indicate the sample dispersion in the range of $5\%$--$95\%$. As a reference, the results for the simulated samples at $z=6$--$9$ are also shown as gray lines \citep[see also Figure 3 of][]{Nakazato:2023}. The red symbols with errorbars are $z>10$ galaxies whose gas-phase metallicity is derived from \OIII 5007\AA\, and H$\alpha$ using MIRI (references are the same as in Figure \ref{fig:Ms_SFR_z10_15}).
}
    \label{fig:MZR_z10_15}
\end{figure}
Figure~\ref{fig:MZR_z10_15} shows the mass--metallicity relation (MZR) at $z=10$--$15$. The gas-phase metallicity is computed from gas cells that host stellar particles younger than 10\,Myr, i.e.\ cells within which such particles are located. This selection is motivated by the fact that observed metallicities are derived from strong-line methods, which rely on
emission lines originating from \HII regions. We find little redshift evolution of the MZR from $z=15$ to $z=10$: both the slope and the normalization remain nearly unchanged. Fitting a linear relation in the $\log M_*$--$\log Z_{\rm gas}$ plane for $z=10-15$ results, we obtain $Z_{\rm gas} \propto M_*^{0.35}$. For comparison, we also show the MZR at $z=6$--$9$, whose slope remains similar up to $M_* \simeq 10^{9.5}\,M_\odot$, above which the relation turns over \citep[see also][]{Nakazato:2023}, consistent with a similar trend recently reported by the COLIBRE simulations \citep{Sharda:2026}. We further plot six $z\geq10$ galaxies observed with MIRI spectroscopy, whose gas-phase
metallicities are derived using the strong-line method based on \OIII 5007\AA\ and H$\alpha$ (references are the same as in Figure~\ref{fig:Ms_SFR_z10_15}). Our simulated galaxies broadly reproduce the observed metallicities. 

Since the dust mass is determined to first order by the product of the gas mass, the gas-phase metallicity, and the dust-to-metal ratio, $M_{\rm d} = {\rm DTM} \times Z_{\rm gas} \times M_{\rm gas}$, we also examine the relation between the stellar mass and the gas mass\footnote{Throughout this work, $M_{\rm gas}$ refers to the gas colder than $10^6$\,K, above which dust cannot survive due to thermal sputtering.}. Although $M_{\rm gas}$ is not directly constrained by observations at $z\geq10$ (even at $z\sim6$--$7$, gas mass estimates suffer from large uncertainties associated with the  \CII-to-gas mass conversion factor), it is a necessary ingredient for interpreting the dust scaling relations presented below. We find $M_{\rm gas} \propto M_*^{0.75}$, with little redshift evolution between $z=15$ and $z=10$. The relation is shown in Appendix~\ref{sec:Ms_Mgas}.

Taken together, the stellar mass--SFR relation and the MZR demonstrate
that our simulations reproduce the key physical properties of observed
$z\geq10$ galaxies over the relevant mass range, providing a realistic
set of gas and stellar mass distributions as input to our dust
radiative transfer calculations.

\begin{figure}
    \centering
    \includegraphics[width =\linewidth, clip]{\figdir/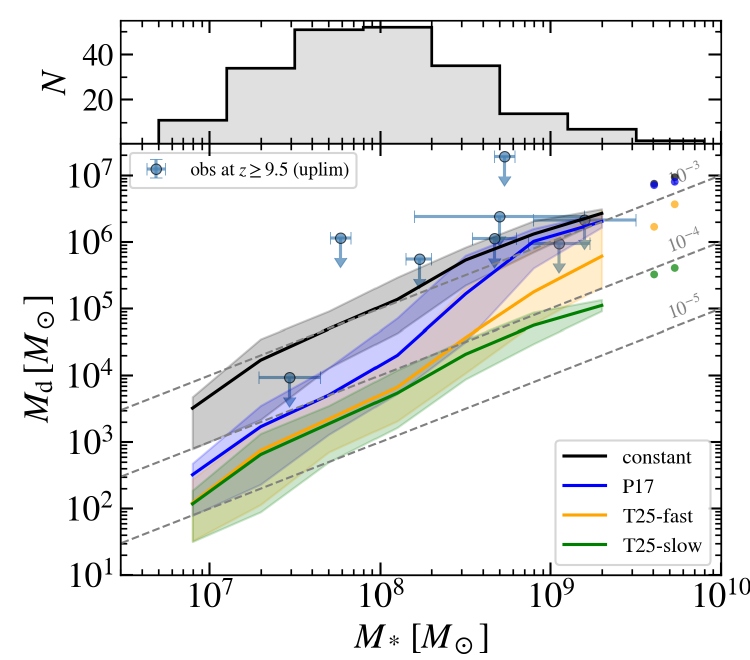}
    \caption{Stellar mass versus dust mass for simulated galaxies from $z=15$ to $z=10$ with different DTM models: \dtmconst\, (black), \dtmP\, (blue), \dtmTfast\, (orange) and \dtmTslow\, (green). The solid lines show the median, and the colored bands indicate the sample dispersion in the range of $5\%$--$95\%$. The blue upper limit symbols are $z\geq9.5$ galaxies with ALMA observations \citep{Bakx:2026, Algera:2026_z8_10_OIII}. 
    The small upper panel shows the stellar mass distribution of our simulated galaxies $z=15$ to $z=10$.
}
    \label{fig:Ms_Md_z10_15}
\end{figure}
\subsection{Dust Mass–Stellar Mass Relation}\label{subsec:dust-to-stellar_mass}
Figure \ref{fig:Ms_Md_z10_15} shows the stellar mass versus dust mass relation for simulated galaxies at $z=10-15$ with different DTM models. We have also examined this relation separately in redshift bins ($z=12.5-15$) and ($z=10-12.5$), and find no significant evolution in the $M_{\rm d}$--$M_*$ relation at $z\geq 10$, justifying our choice to combine all redshifts in Figure \ref{fig:Ms_Md_z10_15}. 

The black line shows the case of \dtmconst\,and we see that it is roughly on the constant dust-to-stellar mass ratio of $M_{\rm d}/M_* = 10^{-3}$. This is because the MZR slope does not change at $z=10-15$ with $Z_{\rm gas} \propto M_*^{0.35}$ as shown in Figure \ref{fig:MZR_z10_15}, and the gas-to-stellar mass ratio is $M_{\rm gas}/M_* \propto M_*^{-0.25}$ in Figure \ref{fig:Ms_Mgas_z10_15} of Appendix \ref{sec:Ms_Mgas}, which together yield $M_{\rm d}/M_* = Z_{\rm gas} \times (M_{\rm gas}/M_*) \propto M_*^{0.1}$, a sufficiently shallow scaling to appear nearly flat over the stellar mass range probed here.

For the \dtmP \, model, $M_{\rm d}/M_* \simeq 10^{-4.5}$--$10^{-4}$ for the low-mass regime with $M_* \lesssim 10^{8.5}\, M_\odot$, and increases and converges to $M_{\rm d}/M_* \simeq 10^{-3}$, similar to the case of \dtmconst. This is because the stellar mass at the turning point has ${\rm log_{10} (O/H) + 12} \simeq 8.0$, corresponding to the critical metallicity as shown in Figure \ref{fig:DTM_Z}.


For the \citetalias{Toyouchi:2025} models, both the fast- and slow-growth cases predict similar dust-to-stellar mass ratios of $M_{\rm d}/M_* \simeq 10^{-5}$--$10^{-4.5}$ in the low-mass regime, $M_* \lesssim 10^{8.5}\,M_\odot$. At higher stellar masses, the rapid-growth model increases to $M_{\rm d}/M_* \simeq 10^{-3.5}$ at $M_* = 10^{9.5}\,M_\odot$ for the same reason as in the \dtmP \, model, whereas the slow-growth model remains nearly constant at $M_{\rm d}/M_* \simeq 10^{-4.5}$.

For comparison, we also plot the current ALMA observational upper limits for $z>10$ galaxies \citep{Bakx:2026}. Around 50\% of the samples are constrained only to $M_{\rm d}/M_* \leq 10^{-3}$, but the other samples suggest that the \dtmconst\, case produces dust masses that are too large. 

\begin{figure*}
    \centering
    \includegraphics[width =0.85\linewidth, clip]{\figdir/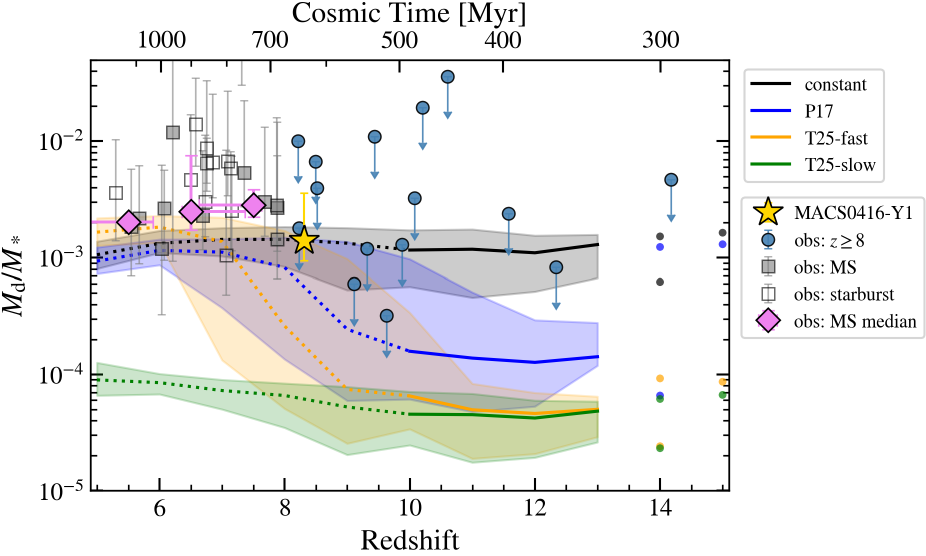}
    \caption{The dust-to-stellar mass ratio as a function of redshift (bottom axis) and cosmic time (top axis) for the simulated galaxies at $z=5$--$15$, with different DTM models as in Figure~\ref{fig:Ms_Md_z10_15}. The solid lines show the median, and the colored bands indicate the $5$--$95$ percentile range of the sample. The dust mass follows directly from the DTM prescription without radiative transfer, so we also show $z<10$, where no RT was performed, as dashed lines. At $z\gtrsim14$, where fewer than three galaxies remain in our sample, we plot individual galaxies instead of the median. The blue symbols with downward arrows are the $3\sigma$ upper limits for $z=8$--$14$ galaxies observed with ALMA \citep{Bakx:2026, Algera:2026_z8_10_OIII}, and the yellow star shows MACS0416-Y1 at $z=8.31$ \citep{Bakx:2025}. At $z<8$ we plot dust-detected galaxies with stellar masses and SFRs updated from JWST observations (references in Appendix~\ref{sec:observational_data}); filled and open grey squares denote main-sequence and starburst galaxies, respectively. The magenta diamonds show the median of the main-sequence galaxies. Note that these are based on detections only and therefore trace
the upper envelope of the underlying population.}
    \label{fig:DtS_redshift}
\end{figure*}

\subsection{Dust-to-Stellar Mass Ratio versus Redshift} \label{subsec:DtS_redshift}
Figure~\ref{fig:DtS_redshift} shows the dust-to-stellar mass ratio as a function of redshift for the different DTM models. Since the dust mass itself can be calculated without performing dust RT calculations, we extend the comparison over $z=5$--15, covering galaxies with stellar masses of $M_*=10^{6.5}$--$10^{11}\,M_\odot$ (see Figure~\ref{fig:Ms_SFR_z10_15}). The median values of $z\geq10$ galaxies are shown by solid lines, while those at $z<10$ are shown by dashed lines. 

For the \dtmconst\ model, $M_{\rm d}/M_*$ shows little redshift evolution because the DTM is fixed and the slope of the MZR changes only weakly over this redshift range (Figure~\ref{fig:MZR_z10_15}). The \dtmTslow\ model similarly predicts an almost constant $M_{\rm d}/M_*$, but at a substantially lower normalization.

In contrast, the \dtmP\ and \dtmTfast\ models show a large scatter in $M_{\rm d}/M_*$, particularly around $z\sim10$. In these models, the DTM depends strongly on metallicity around the critical metallicity for efficient grain growth. Because our simulated sample spans a broad range of stellar masses, and hence gas-phase metallicities through the MZR, galaxies at a given redshift can lie on either side of this transition. This produces a large scatter in DTM and therefore in $M_{\rm d}/M_*$. Toward $z\sim6$, most simulated galaxies have metallicities above the critical value, reducing the scatter. The median $M_{\rm d}/M_*$ consequently increases from $\sim10^{-4}$ at $z\sim13$--10 to $\sim10^{-3}$ at $z\sim6$.

\begin{figure*}
    \centering
    \includegraphics[width=\linewidth, clip]{\figdir/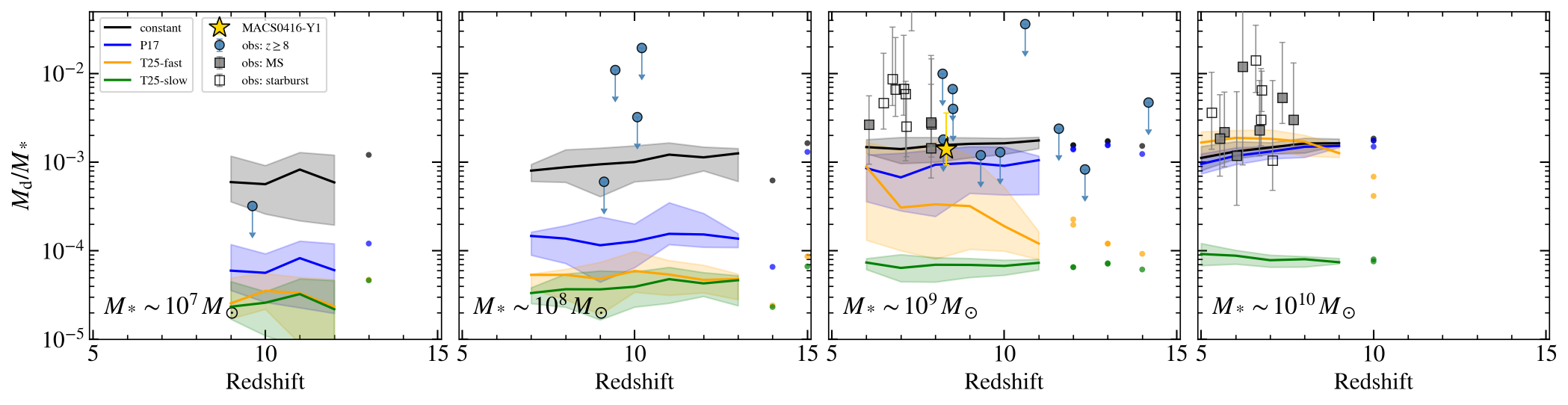}
    \caption{Same as Figure~\ref{fig:DtS_redshift}, but for different stellar mass ranges. The observational results are also plotted in each stellar mass bin. Within each stellar mass bin, the dust-to-stellar mass ratio remains roughly constant across redshift, except for \dtmTfast\ in the middle panel. For the \dtmP\ and \dtmTfast\ models, the galaxies in the two right panels have large $M_{\rm d}/M_*$ values because their metallicities are above the critical metallicity, where dust growth by accretion becomes efficient and increases the dust mass.}
    \label{fig:DtS_redshift_Msbins}
\end{figure*}

To isolate the role of stellar mass, Figure~\ref{fig:DtS_redshift_Msbins} shows the same relation in separate stellar-mass bins. At fixed stellar mass, $M_{\rm d}/M_*$ is nearly independent of redshift in most models. The strong scatter and apparent evolution seen in Figure~\ref{fig:DtS_redshift} therefore arise primarily from the stellar-mass dependence of $M_{\rm d}/M_*$, rather than from strong evolution at fixed $M_*$. This mass dependence is particularly pronounced in the \dtmP\ and \dtmTfast\ models: low-mass galaxies in the left two panels ($M_* \lesssim 10^8 \, M_\odot$) have $M_{\rm d}/M_*\sim10^{-4}$ and $\sim10^{-4.5}$, respectively, whereas galaxies in the right two panels with $M_* \gtrsim 10^9\, M_\odot$ reach values of order $10^{-3}$. This transition reflects the increase in metallicity with stellar mass: galaxies in the higher-mass bins preferentially lie above the critical metallicity, where efficient grain growth rapidly increases the DTM.

The main exception is the \dtmTfast\ model in the intermediate-mass bin, which shows appreciable redshift evolution. Galaxies in this bin evolve across the critical metallicity over the redshift range considered, causing their DTM, and hence $M_{\rm d}/M_*$, to increase rapidly with cosmic time. At metallicities below the critical value, the DTM in the \dtmTfast\ model is a factor of $\sim2$--3 lower than in the \dtmP\ model. Once the critical metallicity is exceeded, however, rapid grain growth raises the DTM above the Milky Way value of 0.4 at $\log({\rm O/H})+12\gtrsim8.5$.

By contrast, the \dtmTslow\ model has too long accretion timescales and an almost metallicity-independent DTM over the relevant metallicity range and therefore predicts $M_{\rm d}/M_*\sim3\times10^{-5}$ with little dependence on either redshift or stellar mass.

\begin{figure*}
\centering
\includegraphics[width=\textwidth]{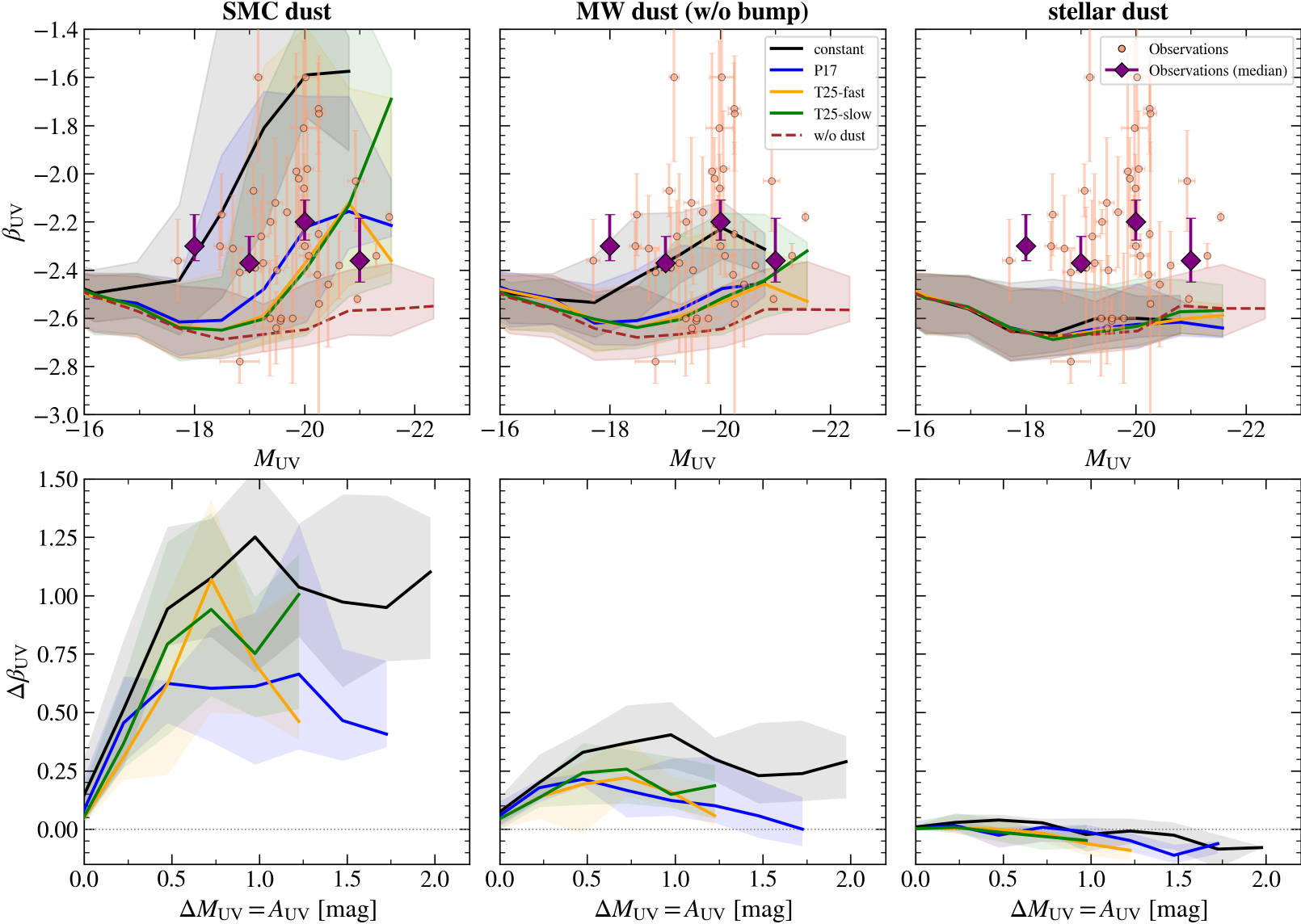}\\
\caption{(Top) UV magnitude versus UV slope for $z=10-15$ galaxies, for SMC-type dust (left), MW-type dust (middle), and stellar dust (right). In each panel, different colors correspond to different DTM models; solid lines show the median and shaded regions the 5th--95th percentile range. For MW dust, $\beta_{\rm UV}$ is measured after masking the UV bump around 2175\,\AA. The brown dashed lines show the intrinsic (dust-free) relation. The orange symbols with error bars are spectroscopically confirmed $z\geq10$ galaxies compiled by \citet{Roberts-Borsani:2026}, and the purple diamonds show their median values in each $M_{\rm UV}$ bin, with errors estimated by bootstrap resampling. (Bottom) The reddening versus the attenuation, $\Delta\beta_{\rm UV}$ versus $\Delta M_{\rm UV}\,(\equiv A_{\rm UV})$, both measured relative to the dust-free case. The stellar dust reddens $\beta_{\rm UV}$ far less than the SMC and MW models at the same $A_{\rm UV}$, and the SMC-like, small-grain dust best covers the observed $\beta_{\rm UV}$ range.}\label{fig:beta_MUV}
\end{figure*}

For comparison, we plot the $z=8$--$14$ upper limits from ALMA \citep{Bakx:2026, Algera:2026_z8_10_OIII},\footnote{Only one galaxy, GN-z11, was observed with NOEMA \citep{Fudamoto:2024}.} together with the highest-redshift dust-detected galaxy MACS0416-Y1 (yellow star, $z=8.31$; \citealt{Bakx:2025}). At $z<8$ we also plot dust-detected galaxies whose stellar masses and SFRs have been updated using JWST observations \citep[][see the compilation in Table \ref{table:obs_data} of Appendix \ref{sec:observational_data}]{Harikane:2020, Witstok:2022, Sommovigo:2022_ALPINE, Algera:2026_REBELS, Valentino:2024, Umehata:2025, Heintz:2025_A1689_zD1, Rajulal:2026}. These are classified using the star-forming main sequence of \citet{Simmonds:2025}: galaxies within $0.5$~dex of the main sequence are shown as filled grey squares, and those above it as open squares. The magenta diamonds show the median of the main-sequence galaxies in bins centered at $z\simeq7.5$, $6.5$, and $5.5$, with error bars indicating the $16$th--$84$th percentiles.\footnote{We exclude REBELS-25, whose stellar mass is likely underestimated and whose dust-to-stellar mass ratio is therefore exceptionally high, $M_{\rm d}/M_*=0.079$ \citep{Algera:2026_REBELS}.}

As discussed in Section~\ref{subsec:dust-to-stellar_mass}, several upper limits at $z\gtrsim9$ disfavor the \dtmconst\ case. The \dtmTslow\ model can also be excluded, since its dust-to-stellar mass ratio remains too low to account for the dust detections at $z\sim5$--$8$ in galaxies of comparable stellar mass ($\log M_*/M_\odot\simeq8.5$--$10$).

Among the dust-detected galaxies at $z<8$, the bursty systems tend to have higher $M_{\rm d}/M_*$ than main-sequence galaxies of comparable stellar mass, and hence of comparable $Z_{\rm gas}$. Since $M_{\rm d}\simeq M_{\rm gas}\cdot{\rm DTM}\cdot Z_{\rm gas}$, 
this offset is naturally attributed to the higher gas fractions associated with elevated sSFR during starburst episodes \citep{Ceverino:2018}. Focusing on the main-sequence galaxies, their $M_{\rm d}/M_*$ values fall within the scatter of the \dtmTfast\ and \dtmP\ models, although the observed medians lie a factor of $\sim2$ above the simulated ones. We caution that the observed dust masses carry two systematic uncertainties. First, the currently dust-detected galaxies represent only the \textit{tip of the iceberg} of the whole population, so the sample is biased towards dust-rich systems. Second, the inferred dust mass depends steeply on the assumed dust temperature as $M_{\rm d}\propto T_{\rm d}^{-(4+\beta_{\rm d})}$. For example, underestimating $T_{\rm d}$ by only $5\,{\rm K}$ (e.g. adopting $40\,{\rm K}$ when the true value is $45\,{\rm K}$) overestimates $M_{\rm d}$ by a factor of $\sim2$. The possibility that the observed $T_{\rm d}$ is underestimated is also pointed out by \citet{Choban:2025}. Indeed, the dust detected $z\simeq 6-7$ galaxies have 0.5-1.0 dex higher
$M_{\rm d}/M_*$ than the local galaxies with the same gas-phase
metallicities \citep{De_Vis:2019} \citep[see Figure 4 of ][]{Algera:2026_REBELS}.

Overall, from the perspective of the dust mass budget, both the $M_*$--$M_{\rm d}$ relation and the redshift evolution of $M_{\rm d}/M_*$ favor metallicity-dependent DTM models, in which efficient grain growth sets in above a critical metallicity.

\begin{figure}
    \centering
    \includegraphics[width =\linewidth, clip]{\figdir/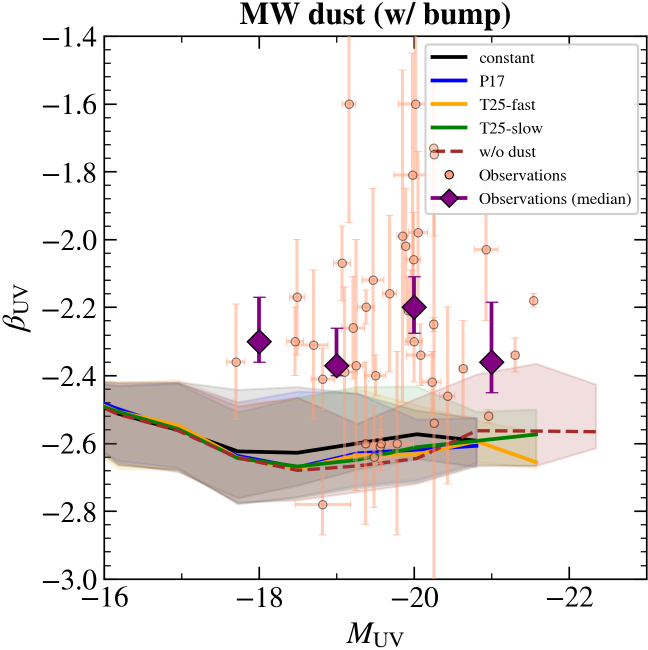}
    \caption{Same as the top panel of Figure~\ref{fig:beta_MUV}, but for MW dust, with the UV slope calculated including the MW bump. Since the bump lies within the wavelength range used to derive the UV slope, it makes the inferred slope bluer, bringing it close to the intrinsic value.}
    \label{fig:beta_MUV_MW_wbump}
\end{figure}

\section{Dust attenuation properties} \label{sec:dust_attenuation_properties}
\subsection{UV Slopes versus UV Magnitude} \label{subsec:UVslope_vs_MUV}
From this section, we present the 3D radiative transfer results at $z=10-15$. The top panels of Figure \ref{fig:beta_MUV} show the relation between the UV slope ($\beta_{\rm UV}$) and the UV magnitude ($M_{\rm UV}$). Here, $\beta_{\rm UV}$ is calculated by fitting the 1300-2200 \AA\ range, following \citet{Roberts-Borsani:2026}. For the MW dust case in the middle panel, we calculate the slope after subtracting the UV bump at 2175 \AA, similar to \citet{Witstok:2023}. This is because, if left uncorrected, the UV bump can bias the inferred slope bluer \citep[e.g.,][]{Narayanan:2018_IRX_beta}. The brown dashed lines in all panels show the intrinsic values (i.e., without dust attenuation, including nebular continuum). The bottom panels show the reddening versus the attenuation relative to the intrinsic values, $\Delta M_{\rm UV}$ versus $\Delta \beta_{\rm UV}$. For comparison, we overplot the spectroscopically confirmed $z\gtrsim10$ galaxies compiled by \citet{Roberts-Borsani:2026}, who measured the UV slope for all 41 galaxies in their sample in the same way from prism spectra.

The left column shows the SMC dust case for the different DTM models. As already discussed in Section \ref{sec:results}, the \dtmconst\ model predicts too much dust at $z \geq 10$. Consequently, galaxies exhibit significant UV reddening even at $M_{\rm UV}\lesssim-16$, resulting in UV slopes that are too red to reproduce the observations. The \dtmP\ and \citetalias{Toyouchi:2025} models (both fast and slow growth), on the other hand, yield similar median values and scatter. Down to $M_{\rm UV} \gtrsim -18.5$, they follow the intrinsic relation ($\beta_{\rm UV, 0} \simeq -2.6$), while for brighter galaxies with $M_{\rm UV} \lesssim -18.5$ the slope becomes redder, with a median of $\beta_{\rm UV} \simeq -2.2$ and a 95th percentile reaching $\beta_{\rm UV} \simeq -1.5$, and the corresponding UV attenuation is $\Delta M_{\rm UV} = A_{\rm UV} = [1.3, 1.7]$. The attenuated $M_{\rm UV}$ values still extend to $M_{\rm UV} \simeq -22$, and the models cover the observed range of the $\beta_{\rm UV}$- $M_{\rm UV}$ relation.

The middle panels show the MW dust case. Down to $M_{\rm UV} \gtrsim -18.5$, the \dtmP\ and \citetalias{Toyouchi:2025} models follow the intrinsic relation, as in the SMC case. At $M_{\rm UV} \lesssim -18.5$, however, the reddening reaches only $\Delta \beta_{\rm UV} \simeq 0.2$, compared with $\Delta \beta_{\rm UV} \simeq 1.0$ for the SMC case. The resulting slopes are therefore bluer than the observed median; although the scatter overlaps with it, the models cannot reproduce the reddest galaxies with $\beta_{\rm UV} > -2$. This indicates that grains smaller than $a \simeq \lambda_{\rm UV}/(2\pi) \simeq 0.025\,\mu{\rm m}$ must dominate in order to explain the range of UV slopes observed at $z > 10$.

The right column shows the stellar dust case. Even for bright galaxies with $M_{\rm UV} \lesssim -18.5$, the reddening in all cases remains small, $\Delta\beta_{\rm UV} < 0.1$, and in most cases the slope becomes even bluer than the intrinsic value. As shown in Figure \ref{fig:size_extinction_SN}, this is because the extinction curve is flat, or even slightly negative, over the UV slope fitting range, and the corresponding attenuation curve might be flat, as we discuss further in Section \ref{subsec:attenuation_curve}. Nevertheless, the range of $M_{\rm UV}$ attenuation itself, $\Delta M_{\rm UV} \sim [1.3, 2.0]$, is comparable to that of the SMC and MW dust cases. Comparing with the observations, all the models lie below the observed median even including their scatter, implying that the typical grain size of $0.1\,\mu{\rm m}$ assumed here is already too {\it blue} to explain the observations.

For the MW case above, we measured $\beta_{\rm UV}$ after subtracting the bump. Slope fitting in observations, however, does not usually correct for the bump\footnote{Except in detailed analyses of the attenuation curve aimed at detecting the bump itself. So far this has been investigated in only a few studies, mainly at $z\sim7$ \citep[e.g.,][]{Witstok:2023, Ormerod:2025}.}. For a fair comparison with the observations, we therefore also calculate the UV slope for MW dust without subtracting the bump. Figure \ref{fig:beta_MUV_MW_wbump} shows the resulting $M_{\rm UV}$ versus $\beta_{\rm UV}$ relation. The bump makes the slope substantially bluer, causing the models to fail to reproduce most of the observations. Since the UV bump spans a few hundred angstroms, even current prism observations of $z \geq 10$ galaxies should be able to detect it if present. This result therefore suggests that most observed $z \geq 10$ galaxies lack the small graphite grains or PAHs that produce the UV bump.

Taken together, the $\beta_{\rm UV}$--$M_{\rm UV}$ relation favors SMC-like dust, characterized by the presence of small grains ($a\lesssim\lambda_{\rm UV}/2\pi\simeq0.025\,\mu{\rm m}$) that contribute to wavelength-dependent UV extinction, together with a low abundance of small graphite grains or PAHs responsible for the 2175 \AA\, bump.

\begin{figure*}
    \centering
    \includegraphics[width =\linewidth, clip]{\figdir/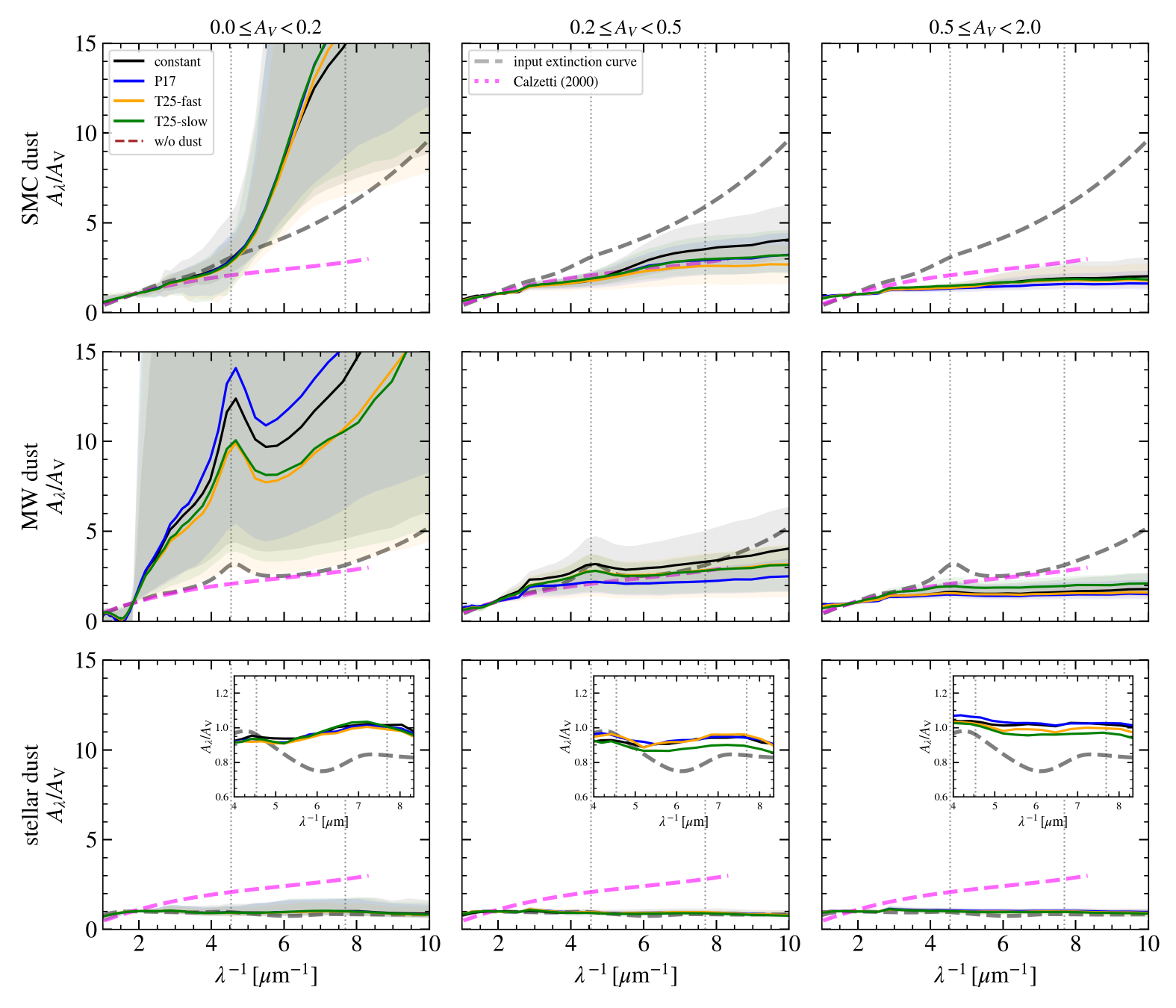}
    \caption{Dust attenuation curves for $z=10-15$ galaxies. Each row shows a different dust model (SMC, MW, and stellar dust from top to bottom), and each column corresponds to a different $A_{\rm V}$ bin. In each panel, different colors indicate different DTM models, with solid lines showing the median and shaded regions the 5th--95th percentile range. For comparison, the pink dashed line shows the Calzetti attenuation curve \citep{Calzetti:2000}, and the grey dashed line shows the input extinction curve adopted for each dust model. For the stellar dust case, the insets show an enlarged view around the UV range. The vertical dotted lines mark the wavelength range used for the $\beta_{\rm UV}$ fit. The SMC and MW attenuation curves start steeper than the input extinction and Calzetti curves but flatten with increasing $A_{\rm V}$, while the stellar dust curves stay flat throughout.}
    \label{fig:attenuation_curve}
\end{figure*}

\begin{figure*}
    \centering
    \includegraphics[width =\linewidth, clip]{\figdir/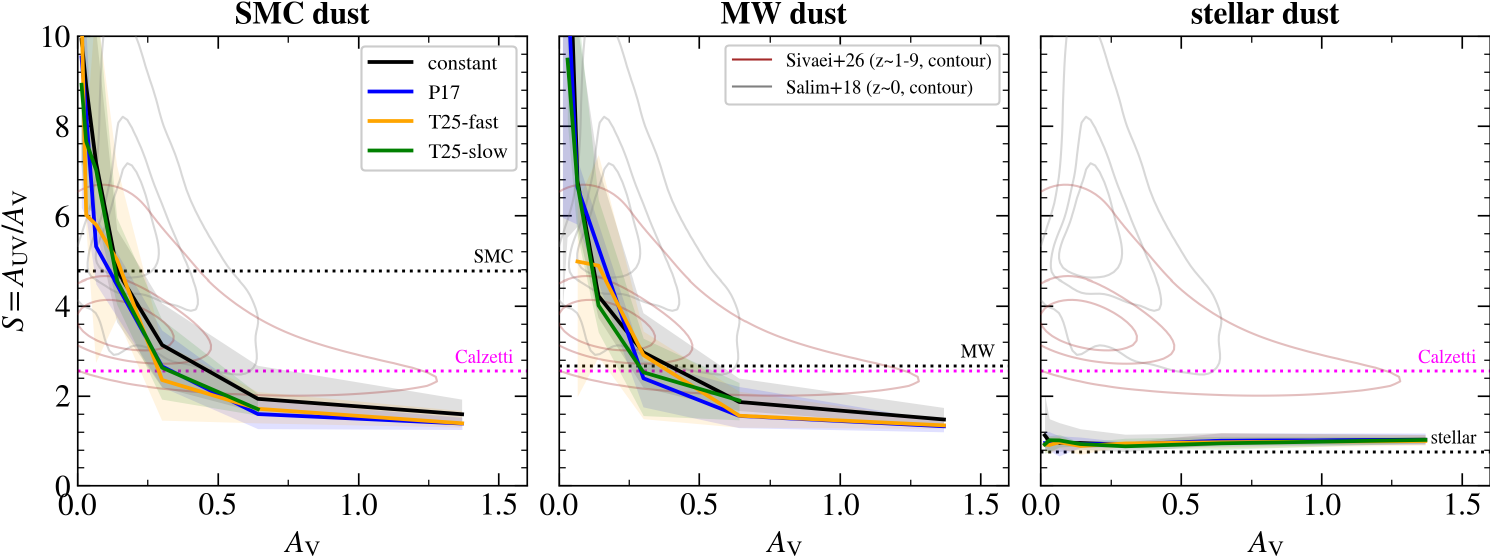}
    \caption{The slope $S = A_{\rm UV}/A_{\rm V}$ as a function of the optical attenuation $A_{\rm V}$ for $z=10-15$ galaxies. Each panel shows a different dust model (SMC, MW, and stellar dust from left to right), with different colors indicating different DTM models, as in Figure~\ref{fig:attenuation_curve}. Solid lines show the median and the shaded regions the 5th--95th percentile range across the simulated galaxy sample. For comparison, the magenta dotted line marks the slope of the Calzetti attenuation curve \citep{Calzetti:2000}, and the black dotted line marks the slope of the input extinction curve adopted for each dust model. For comparison, we also show observational constraints for local star-forming galaxies \citep{Salim:2018} and star-forming galaxies at $z=1$--9 observed with JWST/NIRCam grism spectroscopy \citep{Shivaei:2026}. The observational contours enclose 20\%, 40\%, and 80\% of the estimated two-dimensional KDE probability distribution.
}
    \label{fig:Av_S}
\end{figure*}
\subsection{Attenuation Curves} \label{subsec:attenuation_curve}
Since we obtain both the intrinsic and attenuated SEDs, we can directly calculate the attenuation curve as $A_\lambda = 2.5 \log_{10}(F_{\nu,0}/F_\nu)$, where $F_{\nu,0}$ and $F_\nu$ denote the intrinsic and attenuated flux densities, respectively. Attenuation curves provide important constraints on dust properties, including grain size and composition, as well as the spatial distribution of stars and dust \citep[e.g.,][]{Toyouchi:2026,Narayanan:2018_attenuation_curve,Nakazato:2026,Matsumoto:2026,Sommovigo:2026}. At high redshift, observational constraints on attenuation curves have only recently become available with JWST, extending to $z\sim9$ through NIRCam grism spectroscopy and NIRSpec observations.

Figure \ref{fig:attenuation_curve} shows the dust attenuation curves from the FirstLight simulations at $z=10$--15 for different values of the V-band attenuation, $A_{\rm V}$. From top to bottom, we show the SMC, MW, and stellar dust cases, with different colors indicating different DTM models. For comparison, we plot the Calzetti attenuation curve (pink) and the input extinction curve adopted for each dust model (grey). The vertical dotted lines mark the wavelength range used for the $\beta_{\rm UV}$ fit in Section \ref{subsec:UVslope_vs_MUV}, i.e., 1300--2200 \AA. To quantify the curve slope, we also show the UV-to-V-band attenuation ratio, $S=A_{\rm UV}/A_{\rm V}$, as a function of $A_{\rm V}$ in Figure \ref{fig:Av_S}.
For the SMC dust, all DTM models show very similar attenuation curves, although the \dtmconst\ case is slightly steeper than the other three. At $A_{\rm V}\lesssim 0.2\,{\rm mag}$, all models yield curves steeper ($S>5$) than the input extinction curve ($S=4.8$). This steepening may be partly driven by wavelength-dependent scattering, which can increase the relative contribution of photons scattered into the line of sight at optical wavelengths compared to UV wavelengths \citep[e.g.,][]{Salim_Narayanan:2020,Narayanan:2018_attenuation_curve,Matsumoto:2026,Nakazato:2026}.

As the attenuation increases to $0.2\leq A_{\rm V}<0.5$, the slope decreases to $S=2$--4, approaching the Calzetti curve. At $A_{\rm V}\gtrsim0.5$, the curve becomes shallower than the Calzetti curve. This trend is consistent with previous results for $z=6$--9 FirstLight galaxies assuming MW/SMC dust with \dtmconst\, \citep{Nakazato:2026}. We find that this behavior persists at $z=10$--15 and is largely insensitive to the adopted DTM model.

For the MW dust, all models show steep slopes ($S\gtrsim3$, reaching values above 10) at $A_{\rm V}\lesssim0.2$, together with a prominent 2175 \AA\ UV bump. Since the bump lies within the wavelength range used to measure $\beta_{\rm UV}$, it can significantly affect the measured UV slope. Even excluding the bump, the UV attenuation curve is shallower than in the SMC case. Apart from these differences, the overall behavior is similar to that of the SMC dust: the curves depend weakly on the DTM model, approach the Calzetti curve at $0.2\leq A_{\rm V}<0.5$, and become shallower at higher $A_{\rm V}$. These results suggest that distinguishing MW-like from SMC-like dust may be easier in galaxies with low attenuation ($A_{\rm V}\lesssim 0.2$), where the differences between their attenuation curves are most pronounced. More generally, a flat attenuation curve does not necessarily require a grain size distribution dominated by large grains, but can also arise from radiative transfer effects even for SMC-like dust.

In contrast, the attenuation curves of the stellar dust model differ markedly from those of the SMC and MW dust models. As shown in the right panel of Figure \ref{fig:attenuation_curve}, the input extinction curve of the stellar dust model is nearly grey even in the rest-frame UV. Consequently, wavelength-dependent radiative transfer effects are less pronounced, and the attenuation curve remains nearly flat, with $S\sim1$, over the full range of $A_{\rm V}$.

In Figure \ref{fig:Av_S}, we also compare our results with observational constraints for local ($z\sim0$) star-forming galaxies \citep{Salim:2018} and galaxies at $z=1$--9 \citep{Shivaei:2026}. At fixed $A_{\rm V}$, our simulated $z=10$--15 galaxies generally exhibit shallower attenuation curves than the observational results at lower redshifts. This difference may partly reflect the younger stellar populations in the earliest galaxies. As discussed by \citet{Narayanan:2018_attenuation_curve}, the relative contributions of unobscured young and old stars can significantly affect the attenuation curve: unobscured young stars tend to flatten the curve, whereas unobscured old stars can steepen it. In our $z=10$--15 galaxies, the stellar populations are predominantly young, so even relatively unobscured regions can contribute substantial intrinsic UV emission. UV photons escaping from these regions may limit the increase in the effective UV attenuation as the dust content grows, leading to shallower attenuation curves. In contrast, at lower redshifts, relatively unobscured regions may contain older stellar populations ($>$ 0.5 Gyr) that contribute more strongly to the optical than to the UV emission, resulting in steeper attenuation curves. However, differences in dust properties, star formation histories, and star--dust geometries may also contribute to the observed redshift dependence.

Observational constraints on attenuation curves at $z\gtrsim10$ remain limited, particularly because rest-frame optical wavelengths are shifted beyond the NIRCam wavelength coverage. MIRI spectroscopy can therefore provide important constraints on the attenuation curves of these galaxies. For example, JWST program PRISMS (GO 8051) targets ten spectroscopically confirmed galaxies at $z\sim 10$ with MIRI/LRS, potentially extending attenuation-curve measurements into this redshift regime. Nevertheless, distinguishing SMC-like from MW-like dust may remain challenging because radiative transfer effects can make their attenuation curves similar, particularly at high $A_{\rm V}$ ($A_{\rm V} \gtrsim 0.3$). Lower-attenuation galaxies may offer better opportunities to distinguish these dust models, whereas the nearly grey attenuation predicted for our large-grain-dominated stellar dust model should be more readily distinguishable.

\begin{figure*}
    \centering
    \includegraphics[width =\linewidth, clip]{\figdir/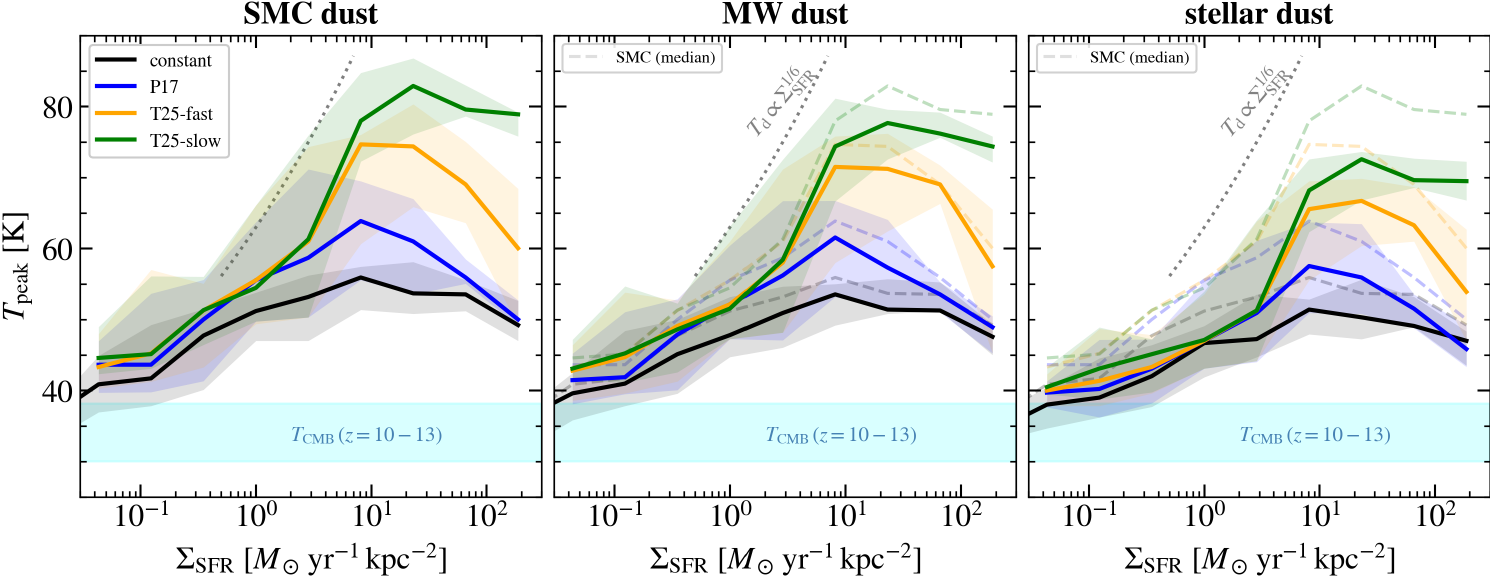}
    \caption{Dust peak temperature versus star formation rate surface density ($\Sigma_{\rm SFR}$) for our simulated galaxies at $z=10-15$ with different DTM models. Each panel shows a different dust model (SMC, MW, and stellar dust from left to right), while the middle and right panels also overlay the median values for the SMC case for comparison. The solid lines show the median values and the shaded regions the 5th--95th percentile range across the simulated galaxy sample. The light blue horizontal band shows the CMB temperature range. The grey dotted line shows the optically thin limit of our analytic model, $T_{\rm d} \propto \Sigma_{\rm SFR}^{1/6}$ (Appendix~\ref{sec:analytical_Td_Sigma_SFR}). $T_{\rm peak}$ depends more on the DTM model than on the grain properties, and it turns over above $\Sigma_{\rm SFR}\sim10\,M_\odot\,{\rm yr^{-1}\,kpc^{-2}}$ for the metallicity-variable DTM models.}\label{fig:Td_Sigma_SFR}
\end{figure*}

\section{Dust reemission properties} \label{sec:dust_reemission_properties}
\subsection{Dust Peak Temperature} \label{subsec:dust_peak_temperature}
In this section and the next, we show the dust re-emission properties. Figure \ref{fig:Td_Sigma_SFR} shows the dust peak temperature as a function of the star formation rate surface density ($\Sigma_{\rm SFR}$) for our simulated galaxies at $z=10-15$, for different DTM models and dust types. We examine this relation because recent semi-analytical calculations suggest that the dust temperature correlates most strongly with $\Sigma_{\rm SFR}$ \citep{Parente:2026_Td}. Here, we define $\Sigma_{\rm SFR} = {\rm SFR}_{10} / (2\pi r_{\rm half, *}^2)$, where $r_{\rm half,*}$ is the projected stellar half-mass radius and ${\rm SFR}_{10}$ is the $10\,{\rm Myr}$-averaged star formation rate. 

For the dust temperature, we calculate the peak dust temperature $T_{\rm peak}$, which is obtained from the Wien displacement law from the rest-frame wavelength at which the infrared flux density ($F_\nu$) peaks ($\lambda_{\rm peak}$)\footnote{The peak flux density is taken as $B_\nu$, not $B_\lambda$.}. This is defined as 
\begin{equation}
T_{\rm d, peak} = \frac{2.898 \times 10^3\, \mu{\rm m}\, {\rm K}}{\lambda_{\rm peak}}, \label{eq:peak_temperature}
\end{equation}
which follows the relation for a graybody with a dust emissivity index $\beta_{\rm d}=2$, i.e., $L_\nu \propto (1-e^{-\tau_\nu}) B_{\nu} \sim \nu^{\beta_{\rm d}}B_\nu = \displaystyle \frac{\nu^{3+\beta_{\rm d}}}{e^{h_{\rm p}\nu/(k_{\rm B}T)}-1}$. This measure has been used in many observational studies \citep[e.g.,][]{Casey:2018, Schreiber:2018, Burnham:2021, Witstok:2022, Mitsuhashi:2024}. Note that $T_{\rm d, peak}$ is simply a proxy for $\lambda_{\rm peak}$, and recent studies suggest that such a single dust temperature does not always recover the true values of dust mass and IR luminosity \citep[e.g.,][]{Sommovigo:2025_Td}. However, we adopt this definition for direct comparison with other theoretical and observational studies.

The left panel shows the results for the SMC dust type. At $\Sigma_{\rm SFR}\lesssim10\,M_{\odot}\,{\rm yr^{-1}\,kpc^{-2}}$, $T_{\rm peak}$ increases with increasing $\Sigma_{\rm SFR}$. This behavior can be understood analytically. In the optically thin limit for UV radiation, the dust temperature scales as
\begin{equation}
T_{\rm d} \propto \Sigma_{\rm SFR}^{1/(4+\beta_{\rm d})}\left(\frac{\kappa_{\rm UV}}{\kappa_{0}}\right)^{1/(4+\beta_{\rm d})} \label{eq:optically_thin_main_text}
\end{equation}
where $\kappa_{\rm UV}$ and $\kappa_0$ are the dust absorption opacities at rest-frame UV and a reference IR wavelength, respectively, and $\beta_{\rm d}$ is the dust emissivity index. This simplified scaling assumes a constant ratio of the stellar and dust half-mass radii; the full derivation is presented in Appendix \ref{sec:analytical_Td_Sigma_SFR}. The gray dotted lines in Figure \ref{fig:Td_Sigma_SFR} indicate the relation $T_{\rm d}\propto\Sigma_{\rm SFR}^{1/6}$ for $\beta_{\rm d}=2$, which qualitatively reproduces the increasing trend of the simulated $T_{\rm peak}$--$\Sigma_{\rm SFR}$ relation. Quantitatively, however, the simulated relations are shallower. Fitting the range $0.3\leq\Sigma_{\rm SFR}/[M_{\odot}\,{\rm yr^{-1}\,kpc^{-2}}]\leq 10$, we obtain slopes of $0.04$--$0.06$ for the \dtmconst\ model and $0.08$--$0.11$ for the other three metallicity variable DTM models, all below the analytical value of 1/6.
These discrepancies may arise from the complex three-dimensional distribution of stars and dust and the resulting range of dust temperatures, which are not captured by our simplified single-temperature analytical model, even in galaxies with globally optically thin dust.
Furthermore, $T_{\rm peak}$ differs by up to $\sim25$ K among DTM models at fixed $\Sigma_{\rm SFR}$, despite the absence of explicit DTM dependence in the optically thin limit, where UV absorption and IR emission both scale linearly with dust mass. This difference suggests that departures from the optically thin limit also play an important role. Indeed, the global UV absorption optical depth reaches unity at $\Sigma_{\rm SFR}\sim 2\,M_\odot\,{\rm yr}^{-1}\,{\rm kpc}^{-2}$ for the \dtmconst\ model, 0.5-0.9 dex lower than for the metallicity-dependent DTM models ($\Sigma_{\rm SFR}\sim 7-13\,M_\odot\,{\rm yr}^{-1}\,{\rm kpc}^{-2}$ for SMC dust; Figure \ref{fig:Sigma_SFR_tau_UV}). This earlier transition likely contributes to the particularly shallow $T_{\rm peak}$--$\Sigma_{\rm SFR}$ relation of the \dtmconst\ model.

At $\Sigma_{\rm SFR}\gtrsim10\,M_{\odot}\,{\rm yr^{-1}\,kpc^{-2}}$, $T_{\rm peak}$ flattens and even decreases slightly with increasing $\Sigma_{\rm SFR}$. This behavior is consistent with the optically thick regime, which is reached at different $\Sigma_{\rm SFR}$ depending on the DTM model, as discussed above. In this limit, the analytical scaling becomes
\begin{equation}
T_{\rm d} \propto \Sigma_{\rm SFR}^{0.048}\left({\rm DTM}\cdot Z_{\rm gas}\right)^{-1/6}\qquad (\beta_{\rm d}=2), \label{eq:optically_thick_main_text}
\end{equation}
where we have assumed a Kennicutt--Schmidt relation $\Sigma_{\rm SFR}\propto\Sigma_{\rm gas}^{1.4}$ to eliminate $\Sigma_{\rm gas}$ (see Appendix \ref{sec:analytical_Td_Sigma_SFR}). The explicit dependence on $\Sigma_{\rm SFR}$ becomes weak, while $T_{\rm d}$ decreases with increasing dust-to-gas ratio as $({\rm DTM}\cdot Z_{\rm gas})^{-1/6}$. In our simulations, ${\rm DTM}\cdot Z_{\rm gas}\propto\Sigma_{\rm SFR}^{\alpha}$, with $\alpha=0.26$--$0.70$ depending on the DTM model, yielding predicted slopes of $+0.004$ to $-0.069$ for the $\log T_{\rm d}$--$\log\Sigma_{\rm SFR}$ relation. These are in reasonable agreement with the measured slopes of $-0.001$ to $-0.094$. The suppression of $T_{\rm d}$ at high dust content can be understood from the heating rate per unit dust mass, $T_{\rm d}\propto(L_{\rm IR}/M_{\rm d})^{1/(4+\beta_{\rm d})}$ for a fixed dust opacity. Once the galaxy becomes optically thick to UV radiation, the absorbed luminosity approaches the available luminosity and becomes relatively insensitive to further increases in dust mass. Consequently, models with lower DTM ratios distribute the absorbed energy over less dust mass, resulting in higher dust temperatures. The weak exponent $-1/(4+\beta_{\rm d})$ reflects the steep temperature dependence of thermal dust emission: for $\beta_{\rm d}=2$, an order-of-magnitude decrease in dust mass increases $T_{\rm d}$ by a factor of $10^{1/6}\simeq1.47$, assuming a fixed absorbed luminosity.

The middle and right panels show the results for the MW and stellar dust types, respectively. For comparison, the median relations for the SMC dust type are overlaid as dashed lines in these panels. The MW case is nearly identical to the SMC case, whereas the stellar dust type yields systematically lower peak temperatures at fixed $\Sigma_{\rm SFR}$ and for the same DTM model across the full range probed.

\begin{figure*}
    \centering
    \includegraphics[width =\linewidth, clip]{\figdir/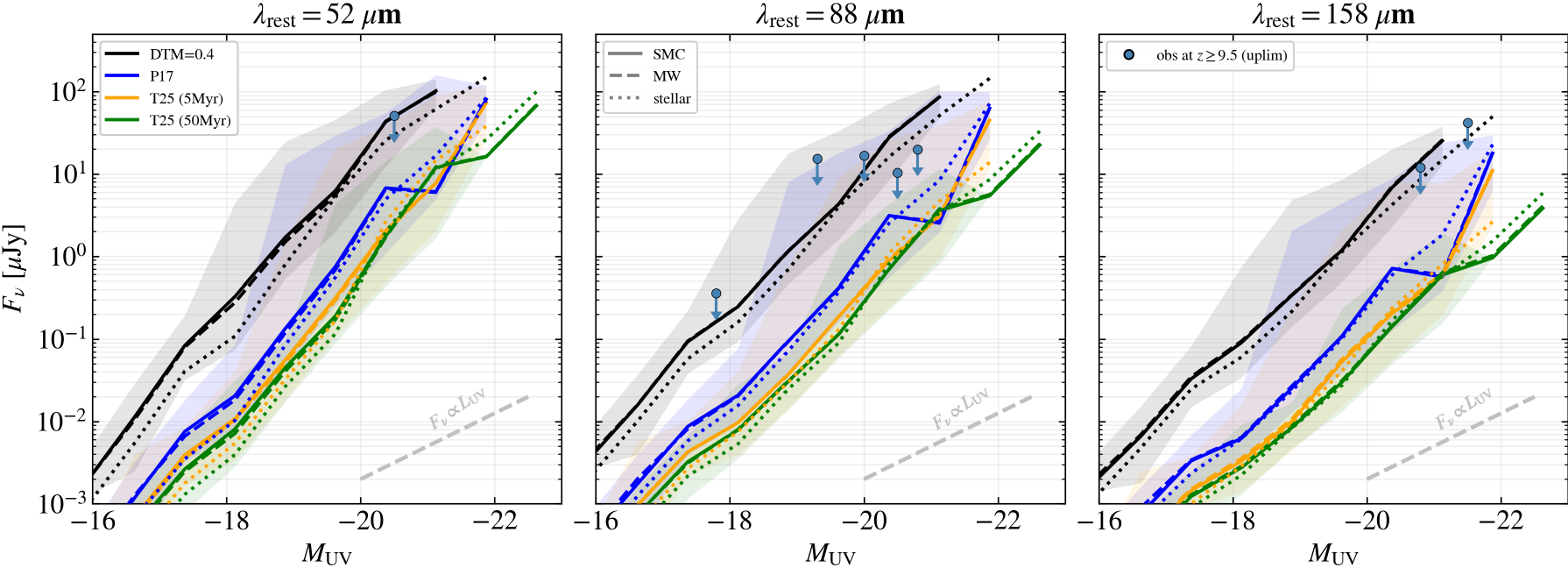}
    \caption{UV magnitude ($M_{\rm UV}$) versus dust continuum flux density at rest-frame wavelengths of $52\,\mu$m (left), $88\,\mu$m (middle), and $158\,\mu$m (right) for our simulated galaxies at $z=10-15$. All fluxes are scaled to the values expected if the galaxies were placed at $z=10$. In each panel, solid, dashed, and dotted lines show the median relations for the SMC, MW, and stellar dust types, respectively, with different colors indicating different DTM models. The shaded regions show the 5th--95th percentile range across the simulated galaxy sample for the SMC dust case. The grey dashed line in each panel indicates the linear relation $F_\nu \propto L_{\rm UV}$. The blue symbols show the $3\sigma$ upper limits for $z\geq9.5$ galaxies from ALMA observations \citep{Bakx:2026, Algera:2026_z8_10_OIII}, with each object plotted in the panel whose wavelength is closest to its observed rest-frame wavelength. Note that several of the observed galaxies are magnified by gravitational lensing, and both the flux densities and the $M_{\rm UV}$ values are shown as de-magnified values. In the most {\it favorable} case (\dtmTfast\, with SMC-like dust), the predicted fluxes lie $\sim1$ dex below the current upper limits, and detections at the same depth would require $M_{\rm UV}\lesssim-21.5$.}
\label{fig:Fnu_MUV_multilambda_demagnified}
\end{figure*}
These differences can be understood in terms of the dust absorption opacities. In the optically thin limit, Eq. (\ref{eq:optically_thin_main_text}) gives $T_{\rm d}\propto(\kappa_{\rm UV}/\kappa_0)^{1/(4+\beta_{\rm d})}$. We obtain $\kappa_{\rm UV}/\kappa_0=3.7\times10^3$, $3.1\times10^3$, and $1.1\times10^3$ for SMC, MW, and stellar dust, respectively, where $\kappa_{\rm UV}$ and $\kappa_0$ are the absorption opacities at 1500 \AA\, and 158 $\mu$m. The stellar dust type has both a lower $\kappa_{\rm UV}$ and a higher $\kappa_0$ than the SMC (and MW) dust type. The former is primarily due to its grain size distribution being dominated by large grains, which have lower UV absorption opacities, while the latter reflects differences in grain composition. Both effects lead to lower equilibrium temperatures. The lower $\kappa_{\rm UV}$ also shifts the transition to globally optically thick conditions towards higher $\Sigma_{\rm SFR}$, by approximately a factor of two in our simulations compared to the SMC and MW cases (Figure \ref{fig:Sigma_SFR_tau_UV}).

In the optically thick limit, the dust temperature becomes insensitive to $\kappa_{\rm UV}$ as the absorbed fraction approaches unity, but retains the dependence $T_{\rm d}\propto\kappa_0^{-1/(4+\beta_{\rm d})}$ (Eq. \ref{eq:Td_optically_thick}). Physically, the temperature is determined by the absorbed luminosity per unit radiating capacity of the dust, $T_{\rm d}^{4+\beta_{\rm d}}\propto L_{\rm IR}/(M_{\rm d}\kappa_0)$. A higher $\kappa_0$ therefore allows the same infrared luminosity to be radiated at a lower temperature. This explains why the stellar dust type remains cooler than the SMC dust type even at the highest $\Sigma_{\rm SFR}$. Assuming the same dust mass, absorbed luminosity, and emissivity index, the expected temperature ratio is $(\kappa_0^{\rm stellar}/\kappa_0^{\rm SMC})^{-1/(4+\beta_{\rm d})}\simeq0.92$, consistent with the temperature difference found in our simulations.

Overall, our results suggest that the choice of DTM prescription has a greater impact on $T_{\rm peak}$ at fixed $\Sigma_{\rm SFR}$ than the dust type, even though the dust models considered here span a wide range of grain sizes and compositions.


\subsection{Rest-frame IR flux density versus $M_{\rm UV}$} \label{subsec:Fnu_vs_MUV}
Figure \ref{fig:Fnu_MUV_multilambda_demagnified} shows the UV magnitude ($M_{\rm UV}$) versus the dust continuum flux density at rest-frame wavelengths of $52\,\mu$m (left), $88\,\mu$m (middle), and $158\,\mu$m (right) for our simulated galaxies at $z=10-15$ with different dust models. We adopt these quantities for both axes because they are directly measurable, which allows us to compare our predictions with observations. The solid lines and shaded regions show the median and the 5th--95th percentile range for the SMC dust, and the dashed and dotted lines show the medians for the MW and stellar dust, respectively. The blue symbols show the $3\sigma$ upper limits for $z \geq 9.5$ galaxies from ALMA observations.

In all panels, the flux density increases rapidly with increasing UV luminosity, spanning more than four orders of magnitude over $-17\gtrsim M_{\rm UV}\gtrsim-22$, and the relations are well described by a single power law over this range. Fitting the median relations at $88\,\mu$m, we obtain $\log F_\nu = -0.86\,M_{\rm UV}+{\rm const}$, or equivalently $F_\nu \propto L_{\rm UV}^{2.15}$, which is steeper than the proportional relation shown by the grey dashed line in each panel. The slope also slightly depends on the wavelength, steepening from $\simeq-0.78$ at $158\,\mu$m to $\simeq-0.86$ at $88\,\mu$m and $\simeq-1.00$ at $52\,\mu$m in all of our models. This is because brighter galaxies have higher $T_{\rm d}$, and the flux density responds more strongly to temperature on the Wien side of the SED than on the Rayleigh--Jeans tail. In all cases, however, the relation is steeper than proportional, $F_\nu \propto L_{\rm UV}^{2-2.5}$.

Comparing the DTM models at fixed dust type, the constant DTM model predicts flux densities about 1 dex higher than those of the other three models at fixed $M_{\rm UV}$, while the \citetalias{Popping:2017} and two \citetalias{Toyouchi:2025} models agree within a factor of a few. This difference directly reflects the predicted dust masses, since $F_\nu \propto M_{\rm d}\,\kappa_\nu\,B_\nu(T_{\rm d})$ in the optically thin limit to IR radiation. As shown in Figure \ref{fig:Ms_Md_z10_15}, the \dtmconst\ model overpredicts the dust mass at $z>10$ by about $1$ dex relative to the other models, and the continuum flux density is overpredicted by a similar amount. All the observational upper limits lie above our predicted median relations for the other three models, indicating that the current non-detections are consistent with our predictions.

Turning to the dust types at fixed DTM model, the differences are much smaller than those between the DTM models. The SMC (solid) and MW (dashed) cases are almost identical, while the stellar-dust case (dotted) gives slightly lower $F_\nu$ over most of the $M_{\rm UV}$ range. Since the UV attenuation is nearly the same among the three dust types (bottom panels of Figure~\ref{fig:beta_MUV}), comparable amounts of stellar radiation are reprocessed into the infrared. The slightly lower $F_\nu$ in the stellar-dust case is therefore mainly due to its slightly larger FIR opacity, $\kappa_0$ (Figure~\ref{fig:kappa_ext_beta_d}), which allows a similar infrared luminosity to be emitted at a lower dust temperature. The differences remain within a factor of two over the entire $M_{\rm UV}$ range, so the dust type cannot be distinguished from the FIR continuum alone.

Finally, we compare the three wavelengths at fixed DTM model and dust type. As discussed in Section \ref{subsec:dust_peak_temperature}, the predicted dust peak temperatures lie in the range $40-80$ K. The temperatures at which the SED peaks at $52$, $88$, and $158\,\mu$m are $55.7$, $32.9$, and $18.3$ K, respectively (assuming $\beta_{\rm d}=2$), so that $52\,\mu$m samples the SED near its peak while $158\,\mu$m lies far out on the Rayleigh--Jeans tail. For galaxies warmer than $T_{\rm d}\simeq40$ K, at which $F_{52}=F_{88}$, the $52\,\mu$m flux density exceeds that at $88\,\mu$m. In practice, however, the rest-frame 52 \mum continuum falls within ALMA Band 8 at $z \simeq 10.5-14.0$, where the sensitivity is lower than in Bands 6 and 7, which cover the rest-frame 88 \mum continuum over this redshift range. Thus, despite the higher predicted continuum flux density at 52 \mum than at 88 $\mu$m, targeting the 88 \mum continuum simultaneously with the \OIII 88 \mum emission line in Band 6 or 7 may be a more practical observational strategy\footnote{At $z \gtrsim 14.4$, however, the 52\mum continuum shifts into Band 7, potentially making it a more favorable target.}. We discuss this strategy further in Section \ref{subsec:dust_detectability}.

\section{Discussion} \label{sec:discussion}
In Sections \ref{sec:dust_attenuation_properties} and \ref{sec:dust_reemission_properties} we have presented the dust attenuation and re-emission properties of our simulated galaxies at $z=10-15$. Here, we discuss the implications of our results for current and future observations. We first examine which combinations of DTM prescription and dust type are consistent with existing observations (Section \ref{subsec:most_favorable_model}), then discuss the detectability of dust continuum emission at these redshifts (Section \ref{subsec:dust_detectability}), and finally address the limitations of our dust models (Section \ref{subsec:caveats}).

\subsection{Favored Combinations of DTM Model and Dust Type} \label{subsec:most_favorable_model}

From the $M_{\rm d}$--$M_*$ relation (Figure \ref{fig:Ms_Md_z10_15}), the metallicity-dependent DTM models (\citetalias{Popping:2017} and \citetalias{Toyouchi:2025}) are favored. Moreover, the redshift evolution of $M_{\rm d}/M_*$ in Figure \ref{fig:DtS_redshift} shows that reproducing both the upper limits at $z>10$ and the dust-detected galaxies at $z=6-8$ requires a DTM ratio that varies with metallicity, with a critical value $Z\sim0.2\,Z_\odot$ above which dust growth via metal accretion becomes effective, as in the \dtmP\ and \dtmTfast\ models. It remains difficult to distinguish between these two. Further constraints on the DTM normalisation in the low-metallicity regime ($Z<Z_{\rm crit}$) would help: as shown in Figure \ref{fig:Ms_Md_z10_15}, the two models could be separated if $M_{\rm d}/M_*>10^{-3.5}$ were constrained at $z>10$. In any case, the metallicity dependence itself is robustly required by our results.

The dust type is constrained by the $M_{\rm UV}$--$\beta_{\rm UV}$ relation. To reproduce the UV slopes of the currently observed $z>10$ galaxies, small grains must already be present in comparable abundance to the $0.1\,\mu$m grains (Figure \ref{fig:size_extinction_SN}); a size distribution dominated by large grains is disfavored. In addition, we find that the UV bump makes the measured slope bluer to $\beta_{\rm UV} \simeq -2.6$, so that dust models producing a strong bump are also disfavored at $z>10$. Taken together, an SMC-like composition, with grains smaller than $a\simeq\lambda_{\rm UV}/2\pi = 0.025\,\mu$m and silicate-dominated in this size regime (i.e., deficit of small graphites with $a < 100$ \AA or PAHs that produce the bump), is the most favorable among the three dust types we consider.

This requirement for small grains is in contrast to  expectations from previous studies. Galaxies at $z>10$ are often assumed to be dominated by SN-driven dust, in which the reverse shock destroys the smallest grains and leaves a size distribution peaked near $0.1\,\mu$m, and the shallow attenuation curves reported at $z=7-9$ have been interpreted as evidence for this picture. We show, however, that although SMC-like dust produces a UV attenuation comparable to that of the large-grain case, it reddens the UV slope enough to match the observations, while its attenuation curve still becomes flat at $A_{\rm V}\gtrsim0.2$. A flat attenuation curve therefore does not by itself require large grains.

Producing SMC-like dust at these redshifts requires either that SN-driven dust is itself dominated by grains smaller than usually assumed, or that $0.1\,\mu$m grains are efficiently processed into smaller ones within a few hundred Myr. A recent analytical calculation by \citet{Toyouchi:2026}, which follows the formation and evolution of dust mass and grain size from $z=20$ to $z=5$, shows that grains smaller than $\sim0.025\,\mu$m can form within $\sim200$~Myr (from $z=19$ to $z=12$) via shattering in the warm neutral medium, supporting the latter possibility.

We do not conclude that dust at $z>10$ is necessarily SMC-like, but our results disfavor an extreme size distribution dominated by large ($\sim0.1\,\mu{\rm m}$) grains, as well as a high abundance of small carbonaceous grains responsible for the 2175\,\AA\ bump, consistent with a silicate-rich composition. However, as discussed in Section~\ref{subsubsec:dust_size_composition}, some galaxies at $z\sim7$ have been detected with UV bumps \citep{Witstok:2023, Ormerod:2025}. Explaining these galaxies requires a mechanism to form small ($<100$\,\AA) carbonaceous grains or PAHs within a few hundred Myr. In this context, we note that on-the-fly dust formation models in isolated-galaxy hydrodynamical simulations have shown that PAHs can form via shattering within a few Myr \citep{Matsumoto:2026}, although this result has been debated \citep{Choban:2026_size}. This suggests that the grain size distribution may vary from galaxy to galaxy, which we leave for future investigation (also Section \ref{subsec:caveats}).

\subsection{Can We Detect Dust Continuum Emission in \(z\gtrsim10\) Galaxies?}
\label{subsec:dust_detectability}
Here we discuss the detectability of the dust continuum at $z\gtrsim10$. As shown in Figure \ref{fig:Fnu_MUV_multilambda_demagnified}, current ALMA observations have searched for the dust continuum mainly in parallel with the rest-frame \OIII 88\,\mum\ line. The depths achieved so far correspond to $3\sigma$ upper limits of $10-20\,\mu$Jy.\footnote{The deepest limits in the figure appear lower only because they have been corrected for lensing magnification; the limits on the observed flux densities are also $10-20\,\mu$Jy.}

As argued in Section \ref{subsec:most_favorable_model}, the most favorable combination among those we consider is a metallicity-dependent DTM model (\dtmP\ or \dtmTfast) with SMC-like dust. We therefore focus on the orange and blue solid curves in Figure \ref{fig:Fnu_MUV_multilambda_demagnified}.

Most of the galaxies spectroscopically confirmed with JWST at these redshifts have $M_{\rm UV}=-18.5$ to $-21$. The brightest is GN-z11 with $M_{\rm UV}=-21.5$ \citep{Tacchella:2023}, which however lies at a northern declination that makes it difficult to observe with ALMA. For a typical target with $M_{\rm UV}=-20$, our favored models predict a rest-frame $88\,\mu$m flux density of $0.5-1\,\mu$Jy, i.e. $10-20$ times fainter than the current depth. Reaching such a depth would require $100-400$ times longer integration times than the $2-20$ hours already invested \citep[see Table 2 of][]{Bakx:2026},\footnote{GN-z11 was instead targeted at rest-frame $\sim158\,\mu$m with NOEMA, for a total of $80.6$ hours \citep{Fudamoto:2024}.} which is prohibitive.

The recently approved ALMA Cycle 12 Large Program PHOENIX (PI: Schouws) will target $15$ high-redshift galaxies across $z=8-15$. Since the metallicity-dependent DTM models predict a rapid increase of $M_{\rm d}/M_*$ around $z\sim9$, the most massive galaxies at $z\lesssim9$ may be detected, although detections at $z\gtrsim10$ remain challenging. Stacking would improve the depth by $\sqrt{N}$, i.e. by a factor of $\sim4$ for $N=15$, reaching $\sim3-5\,\mu$Jy; reaching the predicted $1-2\,\mu$Jy would instead require stacking of order $100$ galaxies.

We also note that gravitational lensing magnifies the UV and the dust continuum by the same factor, so that a lensed galaxy of a given \textit{apparent} $M_{\rm UV}$ is intrinsically fainter, and therefore has a lower observed $F_\nu$, than an unlensed galaxy of the same apparent magnitude. Since $F_\nu\propto L_{\rm UV}^{2}$, lensed UV-bright galaxies are less efficient targets than intrinsically bright ones.

A more promising route is therefore to target intrinsically brighter galaxies. Our models predict that galaxies with $M_{\rm UV}\lesssim-21.5$ should be detectable in the rest-frame $88\,\mu$m continuum at the current depth. Using the double power-law fit to the UVLF at $z\sim12$ from \citet{Harikane:2025_UVLF}, the number density of such bright galaxies is $\lesssim10^{-6.7}\,{\rm cMpc^{-3}}$. Detecting even one such galaxy at $z=12$ would require a survey area of
\begin{equation}
A \simeq \frac{1}{n\,D_{\rm C}^{2}\,({\rm d}D_{\rm C}/{\rm d}z)\,\Delta z}
\simeq 0.96\ {\rm deg^{2}}
\end{equation}
for $\Delta z=1$ and $100\%$ completeness, where $D_{\rm C}$ is the comoving distance and ${\rm d}D_{\rm C}/{\rm d}z = c/H(z)\simeq169.5\,{\rm cMpc}$ at $z=12$. The current JWST/NIRCam surveys (e.g., JADES, COSMOS-Web, CEERS, PRIMER, PANORAMIC) amount to a total of $\sim3000\,{\rm arcmin^{2}}$, or $\sim0.8\,{\rm deg^{2}}$, so that even combining all of them yields an expectation of only $\sim1$ such galaxy. Moreover, most of this area comes from the shallower wide surveys, and whether a galaxy with $M_{\rm UV}=-21.5$ at $z=12$ ($m_{\rm AB}=26.3$) is detectable depends on the depth of each survey.

The upcoming Roman Space Telescope (2027--) will change this situation. Applying the same UVLF to the planned survey areas, we predict that the Roman Ultra Deep, Deep, and High Latitude Surveys will identify $1$, $21$, and $2087$ galaxies at $z\sim12$ with $M_{\rm UV} \lesssim -21.5$, respectively. Such galaxies could then be spectroscopically confirmed with JWST, providing targets for which the dust continuum can finally be detected.

\subsection{Caveats}\label{subsec:caveats}
There are several caveats to our study. First, we adopt four different DTM models: a fixed ratio of $0.4$ ($={\rm DTM_{MW}}$), the metallicity-dependent DTM of \citet{Popping:2017} for their fiducial model, and the metallicity- and redshift-dependent model of \citet{Toyouchi:2025} with two different accretion timescales, $\tau_{\rm acc,0}=5\,{\rm Myr}$ and $50\,{\rm Myr}$. These do not exhaust the range of prescriptions available in the literature: several (semi-)analytical models for the cosmological evolution of the dust mass in high-redshift galaxies at $z\gtrsim6$ have been proposed which we do not consider here \citep[e.g.,][]{Vijayan:2019, Dayal:2022, Tsuna:2023, Parente:2026_size}. In addition, all the DTM prescriptions we adopt depend only on metallicity (and, for \citealt{Toyouchi:2025}, on redshift), and are applied uniformly to gas cells of a given metallicity. Strictly speaking, the timescale of grain growth by metal accretion depends on the local gas properties as $\tau_{\rm acc} \propto n^{-1} T^{-1/2} Z^{-1}$, so that the DTM should also vary with galaxy properties other than metallicity. Just recently, \citet{Narayanan:2026} find that the gas density is the dominant factor controlling dust growth. However, unlike the DTM--metallicity relation, no functional form for the dependence of the DTM on gas density has yet been established, either theoretically or observationally, and we therefore restrict ourselves to metallicity-dependent prescriptions in this work.

Several cosmological simulations also follow the dust mass evolution, albeit typically with a fixed grain size distribution, including DustyGadget \citep{Graziani:2020, Di_Cesare:2023}, CROC \citep{Esmerian_Gnedin:2022, Esmerian_Gnedin:2024}, FIRE \citep{Choban:2025}, DUSTiER \citep{Lewis:2023}, Cosmic Sands \citep{Lower:2023, Lower:2024}, SIMBA-EOR \citep{Jones:2024}, and THESAN \citep{Kannan:2022, Kannan:2025, Garaldi:2026}. A systematic comparison of the rest-IR predictions across these models would be valuable but is beyond the scope of this work.

Second, we consider only three fixed grain size distributions: the SMC and MW types as \textit{vanilla} cases, and the stellar dust type as an extreme case. In reality the grain size distribution evolves with time, and several recent cosmological zoom-in simulations follow this evolution on the fly at $z \gtrsim 6$ \citep[e.g.,][]{Narayanan:2026, Trayford:2026, Vijayan:2026}. Starting from stellar dust with a characteristic size of $0.1\,\mu$m, both \citet{Narayanan:2026} and the COLIBRE simulations \citep{Trayford:2026, Vijayan:2026} find that the small-to-large grain size ratio (STL) increases towards lower redshift, although the amplitude of the increase and its redshift dependence differ between them.

The assumed characteristic size of the stellar dust is itself uncertain. \citet{Toyouchi:2026} have investigated the redshift evolution of the grain size distribution analytically, varying the dust yields and the characteristic size of stellar dust, and adopt $0.01\,\mu$m in their fiducial model, an order of magnitude smaller than the value we assume. This smaller size is motivated by calculations of SN-driven dust destruction by \citet{Otaki:2026}, who find that stellar dust can be eroded down to $a\lesssim0.01\,\mu$m by the reverse shock. Such small grains can grow rapidly to $0.1-1\,\mu$m through metal accretion in cold molecular gas, while shattering in the warm neutral medium simultaneously replenishes the smallest grains ($a\sim10^{-3}\,\mu$m), all within $\sim100\,{\rm Myr}$ (corresponding to the interval between $z=19$ and $z=14$). This allows their model to reproduce the UV luminosity functions at $z\gtrsim10$ and at $z\simeq7$ simultaneously.

In future work, it will be important either to incorporate the redshift evolution of the grain size distribution obtained in the above studies into FirstLight, or to follow the dust evolution in mass and size in post-processing for each gas cell of each snapshot, using the local $n_{\rm gas, cell}, T_{\rm gas, cell}$, and $Z_{\rm gas, cell}$, as done by \citet{Hirashita_Aoyama:2019}. Combined with the rest-IR predictions presented here, such a treatment would allow the UV attenuation and the dust continuum emission to be modelled self-consistently within a single framework.

\section{Summary and conclusions} \label{sec:summary}

The amount and the nature of the dust in the most distant galaxies at $z\gtrsim10$, now routinely discovered with JWST, remain largely unknown. In this study, we have performed post-processed three-dimensional dust radiative transfer calculations on the FirstLight cosmological zoom-in simulations. Our sample consists of 208 snapshots at $z=10-15$, spanning stellar masses of $\log(M_*/M_\odot)=6.5-9.5$ and gas-phase metallicities of $12+\log({\rm O/H})\simeq7.0-8.3$, which covers the range observed for $z\gtrsim10$ galaxies.

We have run the radiative transfer calculations for four different DTM models: a constant ${\rm DTM}=0.4$, the metallicity-dependent DTM of \citet{Popping:2017}, and the metallicity- and redshift-dependent models of \citet{Toyouchi:2025} with fast- and slow-grain growth, combined with three grain size distributions (the SMC, MW, and stellar dust types), giving a total of $208\times12=2496$ radiative transfer runs. Our main results are summarized as follows.

\begin{enumerate}

\item Before performing the radiative transfer, we computed the dust masses predicted by the different DTM models. The \dtmconst\ model maintains a dust-to-stellar mass ratio of $M_{\rm d}/M_*\simeq10^{-3}$ from $z=15$ to $z=6$, exceeding more than half of the upper limits set by the ALMA non-detections at $z\gtrsim10$. In contrast, the \dtmP\ and \dtmTfast\ models predict $M_{\rm d}/M_*\simeq5\times10^{-5}-10^{-4}$ at $z=10-15$, which then increases rapidly at $z\sim8-9$ to reach $\simeq10^{-3}$, as the median gas-phase metallicity approaches $\sim0.2\,Z_\odot$ and grain growth by metal accretion becomes the dominant channel. These two models therefore remain below the observational upper limits at $z\gtrsim10$ while also accounting for the dusty galaxies at the epoch of reionization. The \dtmTslow\ model maintains $M_{\rm d}/M_*\simeq5\times10^{-5}$ throughout $z=15-6$ and cannot reproduce both populations simultaneously.

\item For the attenuation properties, the SMC dust type reddens the UV slope from the intrinsic value of $\beta_{\rm UV, 0}\simeq-2.6$ to a median of $\beta_{\rm UV} \simeq-2.2$ at $M_{\rm UV}\lesssim-18.5$, reaching $\simeq-1.5$ at the 95th percentile, and thus covers the observed range. The MW dust type also reddens the slope, but its median remains bluer than the observed median, while the stellar dust type retains blue slopes over the entire range of $M_{\rm UV}$. This difference is also reflected in the attenuation curves: the stellar dust type yields a flat curve with $S\equiv A_{\rm UV}/A_{\rm V}\simeq1$ across $A_{\rm V}=0-2$ mag, whereas the SMC and MW dust types give curves steeper than that of \citet{Calzetti:2000} at $A_{\rm V}\lesssim0.3$ and flatter at $A_{\rm V}\gtrsim0.3$. In addition, including the 2175 Å bump in the UV slope measurement makes the MW dust model substantially bluer across the entire $M_{\rm UV}$ range. Together, these results favor SMC-like dust containing small grains ($a\lesssim\lambda_{\rm UV}/2\pi\simeq0.025\,\mu{\rm m}$) that contribute to wavelength-dependent UV extinction, along with a low abundance of small graphite grains or PAHs responsible for the UV bump.

\item For the re-emission properties, we find a correlation between the dust peak temperature and $\Sigma_{\rm SFR}$. At $\Sigma_{\rm SFR}\lesssim10\,M_\odot\,{\rm yr^{-1}\,kpc^{-2}}$, the relation approaches the optically thin scaling $T_{\rm d}\propto\Sigma_{\rm SFR}^{1/6}$, while above this value it flattens and follows $T_{\rm d}\propto\Sigma_{\rm SFR}^{0.048}({\rm DTM}\cdot Z_{\rm gas})^{-1/6}$, corresponding to the transition to the optically thick regime for the UV radiation. As a result, $T_{\rm peak}$ becomes sensitive to the DTM model while losing its dependence on $\Sigma_{\rm SFR}$, reaching $60-80\,{\rm K}$ in the starburst regime. These temperatures are systematically higher than the $40-50\,{\rm K}$ typically assumed when converting single-band continuum measurements into dust masses, and the median $\Sigma_{\rm SFR}$ of the spectroscopically confirmed $z\gtrsim10$ galaxies indeed falls in this regime.

\item We also compute the rest-frame IR continuum flux densities as a function of $M_{\rm UV}$. The \dtmP\ and \dtmTfast\ models predict $F_\nu\simeq0.5$- $1\,\mu{\rm Jy}$ at rest-frame $88\,\mu$m for $M_{\rm UV}\simeq -20 \, {\rm mag}$, more than an order of magnitude below the current observational upper limits of $10$—$20\,\mu{\rm Jy}$. At current ALMA sensitivities, continuum detections would instead require UV-brighter galaxies with $M_{\rm UV}\lesssim-21.5 \,{\rm mag}$, which may be identified by upcoming Roman wide-field surveys.

\end{enumerate}
Looking ahead, JWST/MIRI photometry and spectroscopy will provide valuable constraints on the attenuation curves of galaxies at $z\gtrsim10$, helping to disentangle dust grain properties from star--dust geometry and improve estimates of physical properties such as $A_{\rm V}$ and $M_*$. Meanwhile, Roman's wide-field surveys may identify sufficiently UV-bright galaxies at $z\gtrsim10$ within the ALMA-accessible sky. Follow-up ALMA observations of these targets could detect their dust continuum emission at current sensitivities. The combination of wide-area Roman photometry, JWST spectroscopic follow-up, and ALMA continuum observations will thus enable the first direct detections of dust at the cosmic frontier.

\section*{Acknowledgements}
We thank Ikki Mitsuhashi and Kartheik Iyer for fruitful discussions, and Lilian Lee for sharing the CRISTAL observation information. We greatly appreciate Guido Roberts-Borsani, Samir Salim, and Irene Shivaei, who kindly shared their observational data. Numerical computations and analyses were carried out on the Cray XD2000 and analysis servers at the Center for Computational Astrophysics, National Astronomical Observatory of Japan, and on computing facilities at the Flatiron Institute, a division of the Simons Foundation.
YN acknowledges support from the Flatiron Research Fellowship. 
 The authors gratefully acknowledge the Gauss Center for Supercomputing for funding this project by providing computing time on the GCS Supercomputer SuperMUC at Leibniz Supercomputing Centre (Project ID: pr92za). 
The authors acknowledge support by the state of Baden-W\"{u}rttemberg through bwHPC.
     The authors thankfully acknowledge the computer resources at MareNostrum and the technical support provided by the Barcelona Supercomputing Center (RES-AECT-2020-3-0019).
     This work used the v2.1 of the Binary Population and Spectral Synthesis (BPASS) models as last described in \citet{Eldridge:2017}.
     DC is supported by research grant  PID2024-156100NB-C21  and CNS2024-154550 both funded by MI-CIU/AEI/10.13039/501100011033. 
     DT is supported in part by JSPS KAKENHI Grant Number JP25K17438 and the JSPS International Leading Research (ILR) Program (JP22K21349).
     This manuscript benefited from language editing with ChatGPT and Claude. The authors are solely responsible for the scientific content. 
     
\section*{Data Availability}
All data underlying this article, including SEDs and mock images are available upon reasonable request to the corresponding author. 

\appendix 
\counterwithin{figure}{section}
\counterwithin{table}{section}
\section{Nebular continuum contribution} \label{sec:nebular_continuum_contribution}
\begin{figure*}
    \centering
    \gridline{\fig{\figdir/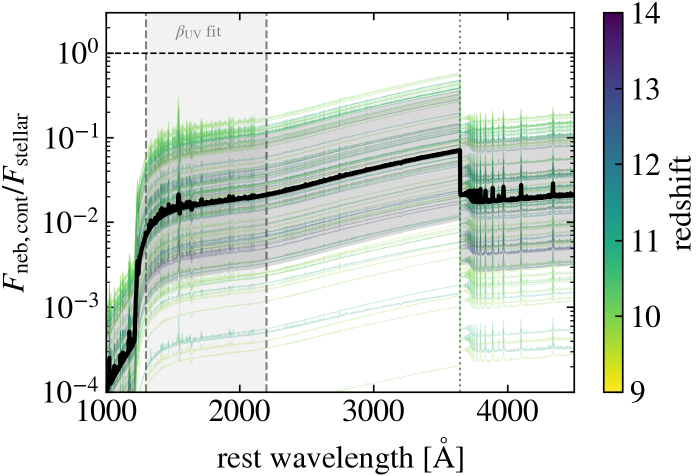}{0.48\linewidth}{}
              \fig{\figdir/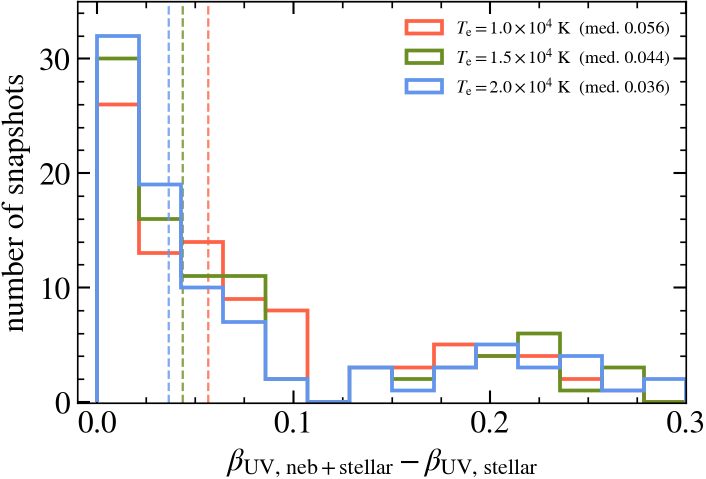}{0.48\linewidth}{}}
    \vspace{-6mm}
    \caption{The contribution of the nebular continuum to the intrinsic SED for our simulated galaxies at $z=10-15$.
\textit{Left:} the ratio of the nebular to the stellar continuum as a function of rest-frame wavelength, color-coded by redshift, assuming an electron temperature of $T_{\rm e}=1.5\times10^{4}\,{\rm K}$. The thick black line shows the median and the grey band the 16th--84th percentile range. The shaded region indicates the wavelength range used to measure $\beta_{\rm UV}$ ($1300-2200\,$\AA; \citealt{Roberts-Borsani:2026}), and the dotted line marks the Balmer jump at $3646\,$\AA.
\textit{Right:} the histogram of the difference in the intrinsic UV slope between SEDs with and without the nebular continuum. The red, green, and blue histograms show the results for $T_{\rm e}/(10^{4}\,{\rm K})=1.0$, $1.5$, and $2.0$, respectively, and the dashed lines mark the corresponding median values.}
    \label{fig:nebular_contribution}
\end{figure*}

\begin{figure}
    \centering
    \includegraphics[width=\linewidth]{\figdir/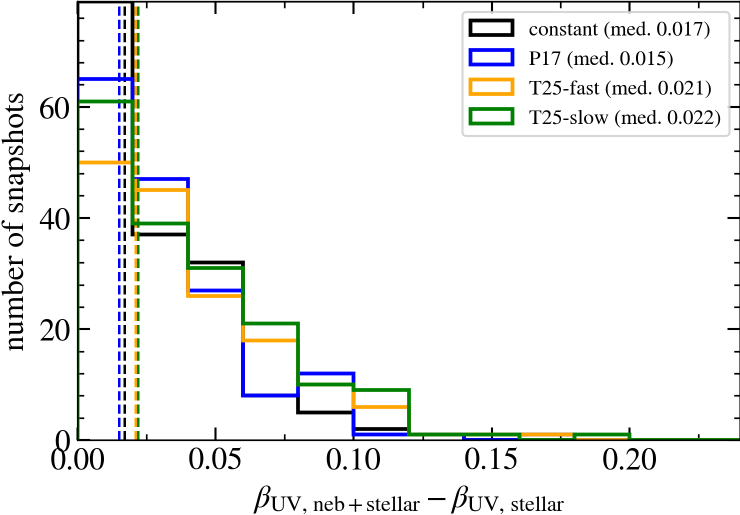}
    \vspace{-6mm}
  \caption{Histogram of the difference in the attenuated UV slope between SEDs with and without the nebular continuum, for the SMC dust type. The nebular continuum is calculated assuming $T_{\rm e}=1.5\times10^{4}\,{\rm K}$. The colors correspond to the different DTM models, and the dashed lines mark the corresponding median values.}
    \label{fig:neb_dbeta}
\end{figure}

In our radiative transfer calculations, we include the nebular continuum in addition to the stellar continuum as the input SED. Here we quantify its contribution to the rest-UV spectrum and its effect on the measured UV slope.

The left panel of Figure \ref{fig:nebular_contribution} shows the ratio of the intrinsic nebular to the stellar continuum, $F_{\rm neb,cont}/F_{\rm stellar}$, for all snapshots, color-coded by redshift. Over the wavelength range used for the $\beta_{\rm UV}$ fit (shaded region), the median contribution is a few per cent, although it reaches $\sim10\%$ in the most actively star-forming systems. To assess the impact on the UV slope, we compute $\beta_{\rm UV}$ over the rest-frame range $1300-2200\,$\AA\ following \citet{Roberts-Borsani:2026}. The right panel of Figure \ref{fig:nebular_contribution} shows the difference in the intrinsic $\beta_{\rm UV,0}$ measured with and without the nebular continuum. The bound-free and free-free emission systematically reddens the UV slope relative to the stellar continuum alone, but the difference remains small: the median offset is $0.044$ for $T_{\rm e}=1.5\times10^{4}\,{\rm K}$, ranging from $0.036$ to $0.056$ for $T_{\rm e}=2.0\times10^{4}$ and $1.0\times10^{4}\,{\rm K}$, respectively.

Figure \ref{fig:neb_dbeta} shows the corresponding change in the UV slope measured after dust attenuation, $\delta \beta_{\rm neb} \equiv \beta_{\rm UV,neb+stellar}-\beta_{\rm UV,stellar}$, for the SMC-like dust. The median shift is $0.015-0.021$ depending on the DTM model, i.e. considerably smaller than the intrinsic offset. This is because the nebular continuum originates from stellar particles younger than $10\,{\rm Myr}$, which are still embedded in dense gas with high dust column densities, so that the nebular component is attenuated more strongly than the older stellar population. Since this shift is much smaller than the reddening caused by dust itself ($\Delta\beta_{\rm UV}\simeq1$ for the SMC dust type; Section \ref{subsec:UVslope_vs_MUV}), our conclusions are not affected by the treatment of the nebular continuum.

\section{Observational data} \label{sec:observational_data}
Here we summarize the references for the observations included in Figure \ref{fig:DtS_redshift} and discussed in Section \ref{subsec:DtS_redshift}: the upper limits on the dust mass for galaxies at $z>8$, and the dust-detected galaxies at $z=5-8$. Table \ref{table:obs_data} lists these galaxies. For the star formation rates we adopt values averaged over $\sim10\,{\rm Myr}$, derived from dust-corrected H$\alpha$ or H$\beta$ where these lines are detected, and from SED fitting otherwise. The gas-phase metallicities are taken from strong-line or direct-method measurements. The tracer used for the SFR is indicated in parentheses in the last column of Table \ref{table:obs_data} where specified in the original reference.

\begin{deluxetable*}{lcccccl}
\tabletypesize{\footnotesize}
\setlength{\tabcolsep}{5pt}
\tablecaption{Observed galaxies included in Figures \ref{fig:DtS_redshift} and \ref{fig:DtS_redshift_Msbins}.}\label{table:obs_data}
\tablehead{
\colhead{ID} & \colhead{$z$} & \colhead{$12+\log({\rm O/H})$} &
\colhead{$\log(M_*/M_\odot)$} & \colhead{SFR $[M_\odot\,{\rm yr^{-1}}]$} &
\colhead{$\log(M_{\rm d}/M_\odot)$} & \colhead{Ref.\ ($Z$; $M_*$; SFR; $M_{\rm d}$)}
}
\startdata
GS-z14-0       & 14.178 & $7.64\pm0.38$          & $8.72^{+0.44}_{-0.40}$  & $25.1^{+5.4}_{-5.6}$   & $<6.38$                & 3; 2; 2; 1 \\
GHZ-2          & 12.333 & $7.40^{+0.52}_{-0.37}$ & $8.27^{+0.23}_{-0.11}$  & $8.7^{+1.4}_{-1.9}$    & $<5.97$                & 5; 4; 4; 1 \\
GS-z11-0       & 11.580 & \nodata                & $8.30\pm0.10$           & $1.45\pm0.10$          & $<6.05$                & --; 6; 6; 1 \\
GN-z11         & 10.603 & $7.91\pm0.07$          & $8.73\pm0.06$           & $24\pm3$               & $<7.28$                & 8; 7; (H$\alpha$) 8; 1 \\
GHZ-1          &  9.875 & \nodata                & $9.20\pm0.30$           & \nodata                & $<6.33$                & --; 1; --; 1 \\
SPT-0615-JD    &  9.625 & \nodata                & $7.14\pm0.18$           & $0.33^{+0.03}_{-0.09}$ & $<3.97$                & 10; 9; 9; 1 \\
GS-z9-0        &  9.433 & $7.49\pm0.11$          & $8.18\pm0.05$           & $5.5\pm0.2$            & $<6.19$                & 11; 11; 11; 1 \\
MACS1149-JD1   &  9.110 & $7.89\pm0.05$          & $7.47\pm0.05$           & $3.2\pm0.3$            & $<4.95$                & 12; 12; (H$\alpha$) 13; 1 \\
ID4590         &  8.496 & $7.16^{+0.10}_{-0.12}$ & $7.15\pm0.15$           & $2.9\pm0.9$            & $<5.59$                & 14; 14; (H$\beta$) 14; 1 \\
MACS0416-Y1    &  8.310 & $7.76\pm0.03$          & $9.00\pm0.07$           & $31.6^{+3.1}_{-2.8}$   & $6.15^{+0.40}_{-0.16}$ & 16; 16; 16; 15 \\
GS-z9-3        &  8.228 & $7.84\pm0.05$          & $9.20^{+0.12}_{-0.13}$  & $13.4\pm1.8$           & $<6.45$                & 18; 17; (H$\alpha$) 18; 1 \\
REBELS-24      &  8.210 & \nodata                & $8.89\pm0.68$           & \nodata                & $<6.90$                & --; 1; --; 1 \\
ZD3ZD6         &  7.881 & $7.94\pm0.05$          & $9.21^{+0.06}_{-0.05}$  & $5.2^{+3.5}_{-1.7}$    & $6.15$                 & 21; 20; 20; 19 \\
YD1            &  7.878 & $8.11^{+0.07}_{-0.05}$ & $8.74\pm0.19$           & $9.1^{+4.3}_{-3.1}$    & $6.17^{+0.71}_{-0.32}$ & 22; 20; 20; 19 \\
ZD12           &  7.876 & $7.44\pm0.05$          & $9.10\pm0.05$           & $3.4^{+2.1}_{-1.1}$    & $6.19$                 & 21; 20; 20; 19 \\
YD4YD6         &  7.874 & $8.22^{+0.11}_{-0.08}$ & $9.25\pm0.11$           & $16.7^{+6.4}_{-5.9}$   & $6.41^{+0.71}_{-0.32}$ & 22; 20; 20; 19 \\
YD7W           &  7.873 & \nodata                & $8.65^{+0.14}_{-0.25}$  & $8.6^{+9.4}_{-4.2}$    & $6.10^{+0.71}_{-0.31}$ & --; 20; 20; 19 \\
REBELS-18      &  7.675 & $8.50\pm0.13$          & $9.71^{+0.06}_{-0.04}$  & $>21$                  & $7.19^{+0.64}_{-0.29}$ & 25; 24; 23; 23 \\
REBELS-12      &  7.346 & $8.23\pm0.13$          & $9.54^{+0.03}_{-0.04}$  & $>26$                  & $7.27^{+0.62}_{-0.29}$ & 25; 24; 23; 23 \\
REBELS-25      &  7.307 & $8.62\pm0.17$          & $9.30^{+0.12}_{-0.14}$  & $>22$                  & $8.20^{+0.60}_{-0.40}$ & 25; 24; 23; 23 \\
SXDF-NB1006-2  &  7.212 & $7.99^{+0.09}_{-0.10}$ & $8.70\pm0.10$           & $62\pm5$               & $<6.59$                & 27; 27; (H$\beta$) 27; 26 \\
B14-65666      &  7.152 & $8.15^{+0.08}_{-0.07}$ & $9.45^{+0.21}_{-0.34}$  & $127\pm66$             & $6.85^{+0.33}_{-0.27}$ & 29; 29; (H$\alpha$) 29; 28 \\
A1689-zD1      &  7.132 & $7.90\pm0.40$          & $9.42^{+0.75}_{-0.11}$  & $137\pm24$             & $7.19^{+0.08}_{-0.10}$ & 31; 30; (H$\beta$+IR) 31; 30 \\
REBELS-14      &  7.084 & $7.90\pm0.12$          & $9.24^{+0.07}_{-0.06}$  & $>45$                  & $7.07^{+0.60}_{-0.29}$ & 25; 24; (H$\beta$) 25; 23 \\
UVISTA-Z-001   &  7.061 & \nodata                & $9.58^{+0.10}_{-0.24}$  & $79^{+142}_{-20}$      & $6.60^{+0.87}_{-0.33}$ & --; 32; 32; 32 \\
REBELS-15      &  6.875 & $7.78\pm0.30$          & $9.31^{+0.02}_{-0.01}$  & $>204$                 & $<7.10$                & 25; 24; 23; 23 \\
COS-3018555981 &  6.854 & \nodata                & \nodata                 & $34^{+10}_{-4}$        & $7.85^{+0.12}_{-0.03}$ & --; --; 32; 32 \\
REBELS-39      &  6.845 & $8.02\pm0.29$          & $9.35^{+0.09}_{-0.08}$  & $>167$                 & $7.17^{+0.58}_{-0.29}$ & 25; 24; 23; 23 \\
COS-2987030247 &  6.808 & $7.54\pm0.15$          & $8.79\pm0.03$           & $32.8\pm8.2$           & $<6.60$                & 33; 33; (H$\alpha$) 33; 32 \\
UVISTA-Z-019   &  6.754 & \nodata                & $9.51^{+0.24}_{-0.15}$  & $63^{+270}_{-27}$      & $7.32^{+0.16}_{-0.03}$ & --; 32; 32; 32 \\
UVISTA-Z-007   &  6.750 & \nodata                & $9.57^{+0.24}_{-0.15}$  & $<128$                 & $<7.00$                & --; 32; 32; 32 \\
REBELS-08      &  6.750 & $8.22\pm0.20$          & $9.33^{+0.07}_{-0.06}$  & $>143$                 & $7.27^{+0.58}_{-0.29}$ & 25; 24; 23; 23 \\
REBELS-32      &  6.729 & $8.48\pm0.13$          & $9.57^{+0.10}_{-0.08}$  & $>104$                 & $7.05^{+0.57}_{-0.28}$ & 25; 24; 23; 23 \\
REBELS-29      &  6.685 & $8.73\pm0.15$          & $9.69^{+0.07}_{-0.05}$  & $>64$                  & $7.05^{+0.56}_{-0.28}$ & 25; 24; 23; 23 \\
REBELS-34      &  6.634 & $8.33\pm0.29$          & $9.45^{+0.03}_{-0.02}$  & $>68$                  & $<7.14$                & 25; 24; 23; 23 \\
REBELS-38      &  6.577 & $8.28\pm0.18$          & $9.75^{+0.09}_{-0.06}$  & $>159$                 & $7.90^{+0.39}_{-0.35}$ & 25; 24; 23; 23 \\
REBELS-05      &  6.496 & $8.51\pm0.16$          & $9.42\pm0.10$           & $>137$                 & $7.09^{+0.55}_{-0.28}$ & 25; 24; 23; 23 \\
J0217-0208     &  6.203 & $8.20^{+0.11}_{-0.15}$ & $10.20^{+0.10}_{-0.20}$ & $159\pm12$             & $8.28^{+1.60}_{-1.10}$ & 27; 27; (H$\beta$) 27; 34 \\
J0235-0532     &  6.089 & $8.20\pm0.11$          & $9.11\pm0.07$           & $142\pm24$             & $<8.84$                & 35; 35; (H$\beta$) 35; 34 \\
Cosmic Grape   &  6.072 & $8.11\pm0.03$          & $8.65^{+0.20}_{-0.12}$  & $2.6^{+1.7}_{-1.5}$    & $6.08^{+0.30}_{-0.40}$ & 37; 37; 37; 36 \\
J1211-0118     &  6.029 & $8.51^{+0.15}_{-0.13}$ & $10.40^{+0.10}_{-0.20}$ & $101\pm9$              & $7.48^{+0.69}_{-0.55}$ & 27; 27; (H$\beta$) 27; 34 \\
DC 417567      &  5.671 & $8.04\pm0.02$          & $9.99\pm0.31$           & $73\pm4$               & $7.33^{+0.33}_{-0.24}$ & 39; 39; 39; 38 \\
DC 683613      &  5.541 & $8.49^{+0.05}_{-0.03}$ & $10.16\pm0.35$          & $40^{+4}_{-5}$         & $7.43^{+0.35}_{-0.24}$ & 39; 39; 39; 39 \\
DC 848185      &  5.294 & $8.28^{+0.03}_{-0.02}$ & $10.30\pm0.28$          & $270^{+24}_{-15}$      & $7.86^{+0.36}_{-0.30}$ & 39; 39; 39; 39 \\
VC 5100994794  &  4.579 & $8.41^{+0.05}_{-0.04}$ & $9.65\pm0.33$           & $78\pm22$              & $6.77^{+0.34}_{-0.23}$ & 39; 39; 39; 39 \\
VC 5101218326  &  4.573 & $8.36^{+0.30}_{-0.34}$ & $10.90\pm0.32$          & $11^{+28}_{-7}$        & $7.40^{+0.34}_{-0.20}$ & 39; 39; 39; 39 \\
DC 818760      &  4.560 & \nodata                & \nodata                 & \nodata                & $8.04^{+0.36}_{-0.28}$ & --; --; --; 39 \\
DC 873756      &  4.546 & $8.68^{+0.88}_{-0.58}$ & $10.53\pm0.08$          & $2^{+8}_{-1}$          & $7.88^{+0.32}_{-0.17}$ & 39; 39; 39; 39 \\
VC 5100822662  &  4.520 & $8.33\pm0.04$          & $10.15\pm0.29$          & $68^{+15}_{-11}$       & $7.20^{+0.35}_{-0.25}$ & 39; 39; 39; 39 \\
VE 530029038   &  4.430 & \nodata                & \nodata                 & \nodata                & $6.97^{+0.35}_{-0.27}$ & --; --; --; 39 \\
\enddata
\tablecomments{References: (1) \citet{Bakx:2026}; (2) \citet{Helton:2026_MIRI1}; (3) \citet{Helton:2026_MIRI2}; (4) \citet{Mitsuhashi:2026_GHZ2}; (5) \citet{Zavala:2025}; (6) \citet{Hainline:2024}; (7) \citet{Bunker:2023}; (8) \citet{Alvarez-Marquez:2025}; (9) \citet{Bradley:2025}; (10) \citet{Messa:2026}; (11) \citet{Curti:2025}; (12) \citet{Marconcini:2024}; (13) \citet{Alvarez-Marquez:2026}; (14) \citet{Heintz:2023_ID4590}; (15) \citet{Bakx:2025}; (16) \citet{Harshan:2024}; (17) \citet{Boyett:2024_DHZ1}; (18) \citet{Bik:2026}; (19) \citet{Umehata:2025}; (20) \citet{Witten:2026_A2744}; (21) \citet{Morishita:2025}; (22) \citet{Venturi:2024}; (23) \citet{Algera:2024_OIII_CII_dust}; (24) Stefanon et al., in prep.; (25) \citet{Rowland:2026_MZR}; (26) \citet{Ren:2023}; (27) \citet{Harikane:2025_ISM}; (28) \citet{Rajulal:2026}; (29) \citet{Prieto-Jimenez:2025}; (30) \citet{Heintz:2025_A1689_zD1}; (31) Blanco-Prieto et al., in prep.; (32) \citet{Witstok:2022}; (33) \citet{Mawatari:2026}; (34) \citet{Harikane:2020}; (35) \citet{Marques-Chaves:2026}; (36) \citet{Valentino:2024}; (37) \citet{Gimenez-Arteaga:2024}; (38) \citet{Sommovigo:2022_ALPINE}; (39) \citet{Faisst:2025}.}
\end{deluxetable*}

\section{Stellar mass versus gas mass}
\label{sec:Ms_Mgas}
\begin{figure}
    \centering
    \includegraphics[width=\linewidth]{\figdir/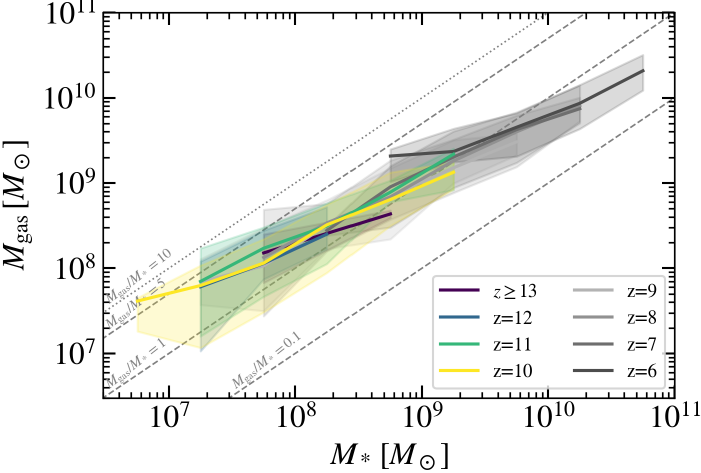}
    \vspace{-6mm}
    \caption{Stellar mass versus gas mass relationship from $z=15$ to $z=10$. Here, $M_{\rm gas}$ is defined as the mass of gas with $T<10^{6}\,{\rm K}$, i.e. the gas in which dust can survive against thermal sputtering. Both stellar mass and gas mass are calculated within a cubic region with a side length of $0.3\,R_{\rm vir}$. The solid lines show the median and the colored bands indicate the sample dispersion in the range of $5\%$--$95\%$. As a reference, the results for the simulated samples at $z=6$--$9$ are also shown as gray lines.
}
    \label{fig:Ms_Mgas_z10_15}
\end{figure}
The gas mass enters our dust mass estimates through $M_{\rm d}={\rm DTM}\times Z_{\rm gas}\times M_{\rm gas}$, and we therefore derive the stellar--gas mass relation of our sample here, even though observational constraints on it remain scarce at these redshifts. Figure \ref{fig:Ms_Mgas_z10_15} shows the relation at $z=10-15$. Here we define $M_{\rm gas}$ as the mass of gas with $T<10^{6}\,{\rm K}$, i.e. the gas in which dust can survive against thermal sputtering.

For stellar masses below $M_*\lesssim10^{8}\,M_\odot$, the gas-to-stellar mass ratio exceeds unity and reaches up to $\simeq5$, corresponding to gas mass fractions of $f_{\rm gas}\equiv M_{\rm gas}/(M_{\rm gas}+M_*)=0.5-0.83$. For the more massive systems comparable to the observed $z>10$ galaxies, $M_*\sim10^{8}-10^{9}\,M_\odot$, the gas mass becomes comparable to the stellar mass. For comparison, we also show further massive systems ($M_*\gtrsim10^{9}\,M_\odot$) at $z=6-9$, where the ratio decreases gradually to $M_{\rm gas}/M_*\simeq0.67$ ($f_{\rm gas}=0.40$) at $M_*\simeq10^{11}\,M_\odot$. Since we find little redshift evolution between $z=15$ and $z=10$, we fit the relation linearly in log-log space and obtain $M_{\rm gas}\propto M_*^{0.75}$.

Observationally, the gas mass can be constrained from the \CII\ line\citep[e.g.,][]{Algera:2026_REBELS}, but no \CII\ detection has yet been reported for galaxies at $z>10$. Moreover, the conversion from $L_{\rm [CII]}$ to gas mass carries a large uncertainty \citep[e.g.,][]{Vallini:2025, Zanella:2018, Heintz:2021}, and it is not clear whether the gas masses derived for \CII-detected galaxies at $z\lesssim8$ correspond to the gas mass defined here. Modelling the \CII\ emission in FirstLight would allow the predicted gas masses to be converted into observable \CII\ luminosities, and hence direct comparison with the observed galaxies; we leave this to future work.

\section{Analytical calculation for $T_{\rm d}$ and $\Sigma_{\rm SFR}$ relationship}
\label{sec:analytical_Td_Sigma_SFR}
Here we analytically derive the dust temperature as a function of the SFR surface density, assuming a single-temperature greybody. In radiative equilibrium, the reemitted IR luminosity equals the absorbed luminosity, $L_{\rm IR} = L_{\rm abs}$.
The IR luminosity can be written as
\begin{align}
L_{\rm IR} &= 4\pi M_{\rm d} \int \kappa_\nu B_\nu(T_{\rm d}) {\rm d}\nu  \notag \\
&= 4\pi M_{\rm d} \kappa_0 \nu_0^{-\beta_{\rm d}}
\int \nu^{\beta_{\rm d}} B_\nu(T_{\rm d}) {\rm d}\nu \notag \\
&= \frac{8\pi k_{\rm B}^{4+\beta_{\rm d}}}
{h^{3+\beta_{\rm d}} c^2 \nu_0^{\beta_{\rm d}}}
\Gamma(4+\beta_{\rm d}) \zeta(4+\beta_{\rm d})
\, M_{\rm d} \kappa_0 T_{\rm d}^{4+\beta_{\rm d}} \notag \\
&\equiv A(\beta_{\rm d}) M_{\rm d} \kappa_0 T_{\rm d}^{4+\beta_{\rm d}} ,
\end{align}
where $\kappa_\nu$ is the mass absorption coefficient per unit dust mass, which is written as $\kappa_\nu = \kappa_0 (\nu/\nu_0)^{\beta_{\rm d}}$ in the Rayleigh limit (i.e., in the rest-IR regime), $\beta_{\rm d}$ is the dust emissivity index,
and $\nu_0$ is a reference frequency at which $\kappa_0$ is evaluated (we adopt $\lambda_0 = c/\nu_0 = 158\,\mu{\rm m}$ throughout). For the conversion from the second to the third line, we use
\begin{equation}
\int_0^\infty \nu^{\beta_{\rm d}} B_\nu(T_{\rm d}) {\rm d}\nu
= \frac{2h}{c^2}\left(\frac{k_{\rm B} T_{\rm d}}{h}\right)^{4+\beta_{\rm d}}
\Gamma(4+\beta_{\rm d}) \zeta(4+\beta_{\rm d}),
\end{equation}
where $h$ is the Planck constant, $c$ is the speed of light, $k_{\rm B}$ is the Boltzmann constant, and $\Gamma$ ($\zeta$) is the Gamma (Riemann zeta) function. Note that $\kappa_0$ and $\nu_0$ enter only through the combination
$\kappa_0 \nu_0^{-\beta_{\rm d}}$, so that $A(\beta_{\rm d})\kappa_0$ is independent of the choice of the reference wavelength.

The absorbed luminosity is written as
\begin{align}
L_{\rm abs} &= f_{\rm abs} L_{\rm bol} \notag \\
&\simeq f_{\rm abs, UV} L_{\rm UV} \notag \\
&= f_{\rm abs, UV} \, \psi_{\rm UV} \, {\rm SFR} .
\end{align}
Here we approximate that the absorbed luminosity is dominated by the UV luminosity, and $\psi_{\rm UV}$ is the UV luminosity per unit star formation rate $[{\rm erg\, s^{-1}/(M_\odot\, yr^{-1})}]$, which depends weakly on the metallicity, the stellar age distribution and the assumed IMF. The absorbed fraction is related to the UV optical depth as
\begin{equation}
f_{\rm abs, UV} = 1 - e^{-\tau_{\rm abs, UV}} , \qquad
\tau_{\rm abs, UV} = \kappa_{\rm UV} \Sigma_{M_{\rm d}} ,
\label{eq:fabs_definition}
\end{equation}
where $\Sigma_{M_{\rm d}}$ is the dust mass surface density and $\kappa_{\rm UV}$ here is the \textit{absorption} opacity at rest-frame UV wavelengths.

Substituting these equations into $L_{\rm IR} = L_{\rm abs}$ and dividing both sides by the surface area, we obtain
\begin{equation}
A(\beta_{\rm d}) \Sigma_{M_{\rm d}} \left(\frac{r_{\rm half, d}}{r_{\rm half, *}}\right)^2 \kappa_0 T_{\rm d}^{4+\beta_{\rm d}}
= f_{\rm abs, UV} \, \psi_{\rm UV} \, \Sigma_{\rm SFR} ,
\label{eq:radiative_equlibrium}
\end{equation}
where $r_{\rm half,*}$ is the half stellar mass radius and $r_{\rm half, d}$ is the half dust mass radius. Equation (\ref{eq:radiative_equlibrium}) together with equation (\ref{eq:fabs_definition}) has two limiting behaviors, which we consider in turn.

In the optically thin limit for UV photons ($\tau_{\rm abs, UV} \ll 1$), we have $f_{\rm abs, UV} \simeq \tau_{\rm abs, UV} = \kappa_{\rm UV} \Sigma_{M_{\rm d}}$, and equation (\ref{eq:radiative_equlibrium}) becomes
\begin{align}
A(\beta_{\rm d}) \Sigma_{M_{\rm d}} \kappa_0 T_{\rm d}^{4 + \beta_{\rm d}} \left(\frac{r_{\rm half, d}}{r_{\rm half, *}}\right)^2
&= \kappa_{\rm UV} \Sigma_{M_{\rm d}} \, \psi_{\rm UV} \, \Sigma_{\rm SFR} .
\end{align}
The dust surface density $\Sigma_{M_{\rm d}}$ cancels out on both sides, because increasing the dust content increases the absorbing and the emitting dust mass in equal proportion, and we obtain
\begin{align}
T_{\rm d} &= \left(\frac{\psi_{\rm UV} \kappa_{\rm UV}}{A(\beta_{\rm d}) \kappa_0}\right)^{1/(4+\beta_{\rm d})}
\left(\frac{r_{\rm half, *}}{r_{\rm half, d}}\right)^{2/(4+\beta_{\rm d})}
\Sigma_{\rm SFR}^{1/(4+\beta_{\rm d})} \label{eq:Td_optically_thin} \\
&\propto \left(\frac{\kappa_{\rm UV}}{\kappa_0}\right)^{1/(4+\beta_{\rm d})} \Sigma_{\rm SFR}^{1/(4+\beta_{\rm d})} .
\end{align}
We have calculated the ratio of the half-dust-mass radius to the half-stellar-mass radius across our simulated sample and found that its
median value remains in the range $\langle r_{\rm half,\star}/r_{\rm half,d}\rangle\simeq0.9$--$2$ for all DTM models and throughout the full $\Sigma_{\rm SFR}$ range probed, so that we can neglect this factor, $\left(r_{\rm half,*}/r_{\rm half, d}\right)^{2/(4+\beta_{\rm d})}\simeq1$. In this regime, the dust temperature is therefore independent of the dust mass and of the DTM ratio, and depends on the dust properties only through the opacity ratio $\kappa_{\rm UV}/\kappa_0$.

In the optically thick limit for UV photons ($\tau_{\rm abs, UV} \gg 1$), the observed UV photons escape from partially optically thin regions, and equation (\ref{eq:radiative_equlibrium}) is rewritten as
\begin{align}
&A(\beta_{\rm d}) \Sigma_{M_{\rm d}} \kappa_0 T_{\rm d}^{4+\beta_{\rm d}} \left(\frac{r_{\rm half, d}}{r_{\rm half, *}}\right)^2
= f_{\rm abs, UV}\psi_{\rm UV} \Sigma_{\rm SFR} \notag \\
\therefore \, & T_{\rm d} = \left(\frac{f_{\rm abs, UV} \psi_{\rm UV}}{A(\beta_{\rm d})\kappa_0}\right)^{1/(4+\beta_{\rm d})} \left(\frac{\Sigma_{\rm SFR}}{\Sigma_{M_{\rm d}}}\right)^{1/(4+\beta_{\rm d})} .
\label{eq:Td_optically_thick_detailed}
\end{align}
Here, we have again omitted the size-ratio term, $\left(r_{\rm half,*}/r_{\rm half,d}\right)^{2/(4+\beta_{\rm d})}\simeq1$,
for the reason given above. We then use $\Sigma_{M_{\rm d}} = \Sigma_{M_{\rm gas}} Z_{\rm gas}
\cdot {\rm DTM}$, and consider the Kennicutt--Schmidt relation $\Sigma_{\rm SFR} = \alpha_{\rm KS} \Sigma_{M_{\rm gas}}^{N}$
\citep{Kennicutt:1998}, where the power-law index is roughly $N\sim 1.4$. We then finally obtain the relation between the dust temperature and $\Sigma_{\rm SFR}$ in the optically thick regime as
\begin{align}
T_{\rm d} &\propto \Sigma_{\rm SFR}^{\left(\frac{1-1/N}{4+\beta_{\rm d}}\right)}
\kappa_0^{-1/(4+\beta_{\rm d})}
({\rm DTM}\cdot Z_{\rm gas})^{-1/(4+\beta_{\rm d})} \notag \\
&\propto \Sigma_{\rm SFR}^{0.048} \, \kappa_0^{-1/6} \,
({\rm DTM}\cdot Z_{\rm gas})^{-1/6} \,\, (\beta_{\rm d}=2, \, N=1.4)\label{eq:Td_optically_thick} .
\end{align}
In contrast to the optically thin case, the dependence on $\Sigma_{\rm SFR}$ is almost negligible, and instead $T_{\rm d}$ decreases with increasing dust-to-gas ratio and with increasing $\kappa_0$. Qualitatively, this is because once the galaxy is optically thick essentially all of the stellar UV radiation is absorbed, so that the heating rate is fixed independently of the dust content; the dust temperature is then set by the heating rate per unit radiating capacity of the dust, $T_{\rm d}^{4+\beta_{\rm d}} \propto L_{\rm IR}/(M_{\rm d}\kappa_0)$, and a fixed energy budget shared among a larger dust mass, or radiated by grains with a higher emissivity, results in a lower temperature.

\begin{figure}
    \centering
    \includegraphics[width=\linewidth]{\figdir/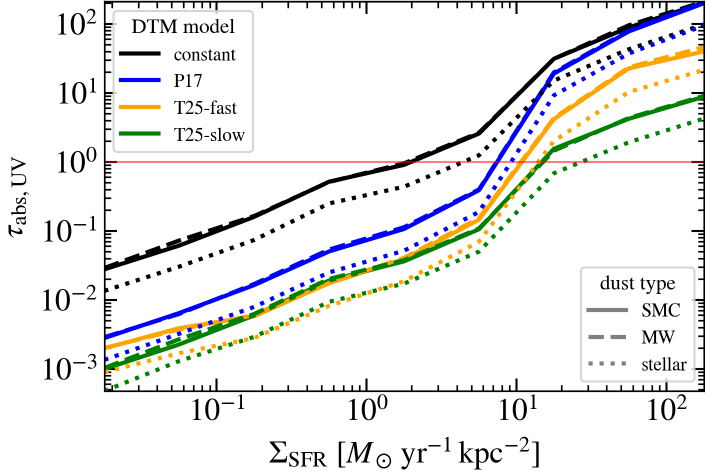}
    \vspace{-6mm}
  \caption{Global UV absorption optical depth at rest-frame 1500 \AA\, as a function of $\Sigma_{\rm SFR}$ for our simulated galaxies at $z=10$–15. Lines show the median values for different DTM models (colors) and dust types (linestyles), as indicated in the legends. The horizontal red line marks $\tau_{\rm abs,UV}=1$. For the metallicity-dependent DTM models, galaxies become optically thick to UV photons at $\Sigma_{\rm SFR}\gtrsim 5$–$10\,M_\odot\,{\rm yr}^{-1}\,{\rm kpc}^{-2}$, whereas the constant-DTM model reaches $\tau_{\rm abs,UV}\gtrsim1$ at $\Sigma_{\rm SFR}\gtrsim 1.5-3\,M_\odot\,{\rm yr}^{-1}\,{\rm kpc}^{-2}$, depending on the dust type.}
    \label{fig:Sigma_SFR_tau_UV}
\end{figure}

To investigate the transition from the optically thin to optically thick regime in our simulated galaxies, we calculate the {\it global} UV absorption optical depth, $\tau_{\rm abs,UV} \equiv \kappa_{\rm UV}\Sigma_{\rm d}$, where $\Sigma_{\rm d} \equiv M_{\rm d}/( 2\pi r_{\rm half,d}^2)$ is the dust surface density and $r_{\rm half,d}$ is the dust half-mass radius. Figure \ref{fig:Sigma_SFR_tau_UV} shows the relation between $\tau_{\rm abs,UV}$ and $\Sigma_{\rm SFR}$ for our simulated galaxies at $z=10$–15. For the three metallicity-dependent DTM models, the median optical depth exceeds unity at $\Sigma_{\rm SFR}\gtrsim 7$–$13\,M_\odot\,{\rm yr}^{-1}\,{\rm kpc}^{-2}$ for the SMC and MW dust models. In contrast, the constant-DTM model reaches $\tau_{\rm abs,UV}\gtrsim 1$ at $\Sigma_{\rm SFR}\gtrsim 1.5\,M_\odot\,{\rm yr}^{-1}\,{\rm kpc}^{-2}$, approximately one order of magnitude lower, owing to its higher dust masses even in less massive galaxies.
The SMC and MW dust models yield nearly identical optical depths because their UV mass absorption coefficients are similar, $\kappa_{\rm SMC,UV}=4.9\times10^4$ and $\kappa_{\rm MW,UV}=5.1\times10^4\,{\rm cm}^2\,{\rm g}^{-1}$. In comparison, the stellar dust model has a lower opacity, $\kappa_{\rm stellar,UV}=2.4\times10^4\,{\rm cm}^2\,{\rm g}^{-1}$, resulting in smaller optical depths and shifting the transition to somewhat higher $\Sigma_{\rm SFR}$. Overall, the transition from optically thin to optically thick conditions depends more strongly on the adopted DTM prescription than on the dust type.

\bibliography{FirstLight_dust_z10}{}
\bibliographystyle{aasjournal}

\end{document}